\documentclass[1p]{elsarticle}
\usepackage{amsmath,amssymb}
\usepackage{graphicx}
\def\vec#1{{\boldsymbol#1}}                % bold vector
\def\vev#1{{\langle#1\rangle}}             % expectation value
\def\abs#1{\left|#1\right|}                % absolute value
\let\Re\relax\DeclareMathOperator{\Re}{Re} % real part
\let\Im\relax\DeclareMathOperator{\Im}{Im} % imag. part
\DeclareMathOperator{\erf}{erf}            % error function
\DeclareMathOperator{\tr}{tr}              % trace
\def\k{{\vec k}}                           % commonly used vectors...
\def\p{{\vec p}}
\def\q{{\vec q}}
\def\r{{\vec r}}
\def\x{{\vec x}}
\def\SE{{\text{S}}}                        % self-energy
\def\MT{{\text{MT}}}                       % Maki-Thompson
\def\AL{{\text{AL}}}                       % Aslamazov-Larkin
\def\ret{{\text{ret}}}                     % retarded

\begin{document}
\begin{frontmatter}
\title{Viscosity and scale invariance in the unitary Fermi gas}
\author[m]{Tilman Enss}
\author[k]{Rudolf Haussmann}
\author[m]{Wilhelm Zwerger}
\address[m]{Technische Universit\"at M\"unchen,
  James-Franck-Stra\ss{}e, D-85747 Garching, Germany}
\address[k]{Fachbereich Physik, Universit\"at Konstanz,
  D-78457 Konstanz, Germany}

\begin{abstract}
  We compute the shear viscosity of the unitary Fermi gas above the
  superfluid transition temperature, using a diagrammatic technique
  that starts from the exact Kubo formula. The formalism obeys a Ward
  identity associated with scale invariance which guarantees that the
  bulk viscosity vanishes identically. For the shear viscosity, vertex
  corrections and the associated Aslamazov-Larkin contributions are
  shown to be crucial to reproduce the full Boltzmann equation result
  in the high-temperature, low fugacity limit. The frequency dependent
  shear viscosity $\eta(\omega)$ exhibits a Drude-like transport peak
  and a power-law tail at large frequencies which is proportional to
  the Tan contact.  The weight in the transport peak is given by the
  equilibrium pressure, in agreement with a sum rule due to Taylor and
  Randeria.  Near the superfluid transition the peak width is of the
  order of $0.5\, T_F$, thus invalidating a quasiparticle description.
  The ratio $\eta/s$ between the static shear viscosity and the entropy density
  exhibits a minimum near the superfluid transition temperature whose
  value is larger than the string theory bound $\hbar/(4\pi k_B)$ by a
  factor of about seven.\\[1ex]
  PACS: 67.10.Jn; 67.85.Lm; 11.30.-j
\end{abstract}
\end{frontmatter}

\section{Introduction}
\label{sec:intro}

The remarkable derivation of a simple proportionality between the
shear viscosity $\eta$ and the entropy per volume $s$ in a $\mathcal
N=4$ supersymmetric Yang-Mills theory in the limit of infinite 't
Hooft coupling $\lambda=g^2N\to\infty$ by Policastro, Son and
Starinets \cite{policastro2001shear} and the conjecture by Kovtun, Son
and Starinets (KSS) \cite{kovtun2005vsi} that the ratio $\eta/s$ is
larger than the value $\hbar/(4\pi k_B)$ found in this limit for {\it
  all} scale invariant, relativistic field theories have motivated the
search for the `perfect fluid' which realizes, or at least comes
close to, this bound \cite{schaefer2009npf}.  In spite of some
theoretical counter-examples \cite{cohen2007mpf, son2008comment,
  brigante2008viscosity, kats2009effect, buchel2009beyond}, the KSS
conjecture turns out to be valid for 
all real fluids that are known.  In particular, since the velocity of light
does not appear in the KSS bound, the conjecture is applicable to both
relativistic and non-relativistic field theories and may even be
extended to complex, classical fluids.  In the case of water, for
instance, the minimum value for the ratio $\eta/s$ is found close to
its critical point at $650\,$K and is only a factor of $25$ above the
KSS bound \cite{kovtun2005vsi, schaefer2009npf}.
Clearly, $\hbar$ is irrelevant for the minimum value of $\eta/s$ of
water at these temperatures. Yet, as shown in \ref{sec:LJ},
there is a simple argument which shows that the viscosity minimum of
purely classical fluids is not far above that expected from the KSS
bound.  This begs the question what are the necessary conditions for a
fluid to be `perfect' in the sense of a minimum value of $\eta/s$ that
is limited only by quantum mechanics and, moreover, what is the role
of scale invariance in this context?  As pointed out by Kovtun, Son
and Starinets \cite{kovtun2005vsi} and discussed in more detail in a
number of recent lecture notes~\cite{hartnoll2009lectures,
  mcgreevy2009holographic, sachdev2010condensed} on connections
between holographic duality and many-body physics, a crucial
requirement for a fluid to come close to the KSS bound is the fact
that it is strongly interacting and thus has no well-defined
quasiparticles.  Indeed, in a situation with proper quasiparticles,
transport coefficients like the shear viscosity can be computed using
kinetic theory. Since the lifetime broadening $\hbar/\tau$ is much
less than the average energy $k_BT$ for well-defined quasiparticles,
the resulting $\eta/s \gg \hbar/k_B$ is typically far above the KSS
bound \cite{schaefer2009npf}.  By contrast, for strongly coupled
quantum field theories, the relaxation times are expected to be of
order $\hbar/(k_BT)$. Since $\eta/s T$ is a characteristic time scale
for shear relaxation, this immediately implies that $\eta/s$ is of
order $\hbar/k_B$ in the strongly coupled limit. A nontrivial example
in this context is the standard $SU(3)$ Yang-Mills theory. The
associated $\eta/s$ ratio for the pure gauge theory has been
calculated numerically using lattice QCD~\cite{meyer2007cot,
  meyer2008cbv, meyer2008energy}. Its minimum appears near the
deconfinement transition temperature and turns out to be rather close
to the KSS bound. The standard Yang-Mills theory near $T_c$ and its
supersymmetric extension---for which $\eta/s$ is independent of
temperature and no well-defined quasiparticles exist at arbitrary
energies---thus have very similar values of $\eta/s$. Adding fermions
to the pure gauge theory, the ratio $\eta/s$ becomes an experimentally
accessible quantity in high-energy, non-central collisions of heavy
nuclei.  The observed ratio $\eta/s$ of the quark-gluon plasma is
around $0.4\,\hbar/k_B$, i.e., a factor five above the KSS
bound~\cite{schaefer2009npf}.

In a condensed matter context, generic examples for strongly coupled,
finite temperature field theories for which no quasiparticle
description holds are provided by models that exhibit a zero
temperature critical point \cite{sachdev1998quantum}.  These systems
are scale invariant at a critical value $g_c$ of the coupling $g$.
At finite temperature $T$, there is a quantum critical regime above
the critical point, which covers a finite window $T>|g-g_c|^{z\nu}$.
In this regime, the thermal energy $k_BT$ is the only energy scale and
correlations of the order parameter exhibit incoherent relaxation with
a characteristic time scale $\tau_{\Psi}=\mathcal C\,\hbar/(k_BT)$
\cite{sachdev2010condensed}.  Here, $\mathcal C$ is a 
constant that only depends on the universality class of the quantum
phase transition.  In addition, universal behavior shows up in
transport coefficients like the conductivity and also the shear
viscosity in the hydrodynamic regime $\hbar\omega\ll k_BT$. A concrete
example is the pseudo-relativistic theory of graphene where the ratio
$\eta/s=\Phi_{\eta}\hbar/k_B$ has recently been calculated within a
Boltzmann equation approach. The marginally irrelevant Coulomb
interaction in this case gives rise to a logarithmic temperature
dependence $\Phi_{\eta}=0.008\ln^2(1/T)$ of the prefactor
$\Phi_{\eta}$ \cite{mueller2009graphene}.  This yields a monotonically
increasing viscosity as the temperature approaches zero within
the quantum critical regime. Logarithmic singularities as $T\to 0$
are also present in standard 2d Fermi liquids \cite{novikov2006viscosity}.

Here, we consider the shear viscosity for the unitary Fermi gas, a
system of attractively interacting Fermions at infinite scattering
length.  The unitary Fermi gas is realized experimentally with
ultracold atoms in a balanced mixture of, e.g., the two lowest
hyperfine levels of $^6$Li at a Feshbach resonance and has been
studied quite extensively over the past few years
\cite{ketterle2008mpa, bloch2008mbp, giorgini2008tuf}. It provides an
example of a non-relativistic field theory that is both scale- and
conformally invariant \cite{nishida2007ncf}.  The underlying quantum
critical point in this case is the zero density gas at unitarity, as
was shown by Nikoli\'c and Sachdev \cite{nikolic2007rg}.  As a
consequence of scale invariance, pressure $p$ and energy density
$\epsilon$ of the gas are related by $p=2\epsilon/3$
\cite{ho2004universal}. Moreover, the bulk viscosity vanishes at all
temperatures \cite{son2007vbv, werner2006unitary}. For the shear
viscosity, quantitative results so far are only available in the
high-temperature limit $T\gg T_F$ where a classical description in
terms of a Boltzmann equation is possible \cite{bruun2007sva}, and
also deep in the superfluid regime $T\ll T_c\simeq 0.15\, T_F$. In the
superfluid, a finite viscosity arises from phonon-phonon collisions in
the normal fluid component, giving rise to a rapid increase
$\eta(T)\sim T^{-5}$ of the viscosity as the temperature approaches
zero \cite{rupak2007svs}.  Since $\eta(T)\sim T^{3/2}$ also increases
in the classical limit, both the viscosity and the ratio $\eta/s$ of
the unitary gas necessarily exhibit a minimum, a behavior which is in
fact typical for any fluid \cite{schaefer2009npf}. The major question
which will be discussed in this paper, is where this minimum appears
and what the associated value of the viscosity and entropy is. It
turns out that the minimum in $\eta/s$ for the non-relativistic
unitary gas is about a factor of seven above the KSS bound, rather
close to the value that is found for the relativistic quark-gluon
plasma.  We also show that the minimum in $\eta$ implies a lower bound
on the shear diffusion constant $D_{\eta}$, which is about
$0.5\,\hbar/m$.

In detail, we determine the real part $\eta(\omega)$ of the
frequency-dependent shear viscosity from a diagrammatic method that
evaluates the stress tensor correlation function in the exact Kubo
formula.  Within a conserving approximation that respects all
symmetries and the associated conservation laws, we obtain
$\eta(\omega)$ in the normal phase and, in particular, the static
viscosity to entropy density ratio $\eta(\omega=0)/s$.  Our results
for $\eta(\omega)$ show a Drude-like transport peak around $\omega=0$.
Its width defines a viscous scattering rate $1/\tau_\eta$ which obeys
$\hbar/\tau_{\eta}\ll k_BT$ at high temperatures, where kinetic theory
is applicable. Near the transition temperature $T_c\simeq 0.15\, T_F$
to the superfluid phase of the unitary gas, the width approaches
$0.5\,k_BT_F$, thus clearly invalidating a quasiparticle description
of viscous transport in this regime.  The weight $W=2\epsilon/3=p$ of
the Drude peak turns out to be equal to the pressure at all
temperatures, consistent with a sum rule derived recently by Taylor
and Randeria \cite{taylor2010viscosity}.  For large frequencies
$\hbar\omega \gtrsim k_BT$, there is a crossover from the Drude peak
to an inverse square-root tail $\eta(\omega) \sim 1/\sqrt\omega$,
whose amplitude is proportional to the Tan contact density $C$
\cite{tan2008energetics}. 
In the high-temperature limit $T\gg T_F$ we
complement our numerical solution of the transport integral equations
by an analytical solution to leading order in the fugacity
$z=e^{\beta\mu}$.  We thus confirm that $(i)$ the Kubo formula yields
exactly the same expression as the Boltzmann equation in this limit,
and $(ii)$ the vertex corrections, in particular the Aslamazov-Larkin
contributions, are crucial even in the classical limit and increase
the scattering time by a factor of almost three.  This resolves an
inconsistency between the Boltzmann equation result and a previous
diagrammatic calculation in the high-temperature limit by Bruun and
Smith \cite{bruun2007sva}.  Finally, we derive Ward identities
associated with the scale invariance of the unitary gas and show that
they are obeyed within our approximation for the bulk viscosity
vertices at all frequencies and temperatures.  As a consequence, the
bulk viscosity vanishes identically $\zeta(\omega) \equiv 0$
\cite{son2007vbv, nishida2007ncf, taylor2010viscosity}.

The paper is organized as follows: in section \ref{sec:unitary} we
define the microscopic model for the unitary gas and discuss its scale
invariance.  The basic correlation functions that allow to calculate
the shear viscosity from first principles via a Kubo formula are
defined in section \ref{sec:kubo} together with a discussion of
hydrodynamic relations and sum rules.  In section \ref{sec:alpha} we
give a qualitative discussion of the temperature dependence of the
shear viscosity in the classical and in the deep superfluid limit, as
well as near the superfluid transition temperature.  The transport
integral equations and their solution within a self-consistent
Luttinger-Ward formalism are derived in section \ref{sec:calc}.  An
exact solution of these equations to leading order in the fugacity is
presented in section \ref{sec:boltzmann}, which turns out to reproduce
the full solution of the Boltzmann equation in the high temperature
limit.  In section \ref{sec:results} we discuss our results for the
frequency dependent viscosity in the low temperature---but still
normal fluid---regime of the unitary gas, where the assumption of
independent collisions of well defined quasiparticles that underlies
the Boltzmann equation breaks down.  In particular, we calculate the
shear viscosity to entropy density ratio $\eta/s$, which displays a
minimum value slightly above the transition to the superfluid.
Conclusions and open questions are presented in section
\ref{sec:concl}.  There are three appendices: in
\ref{sec:LJ}, we give an argument based on purely dimensional
analysis which explains why the inherently quantum mechanical KSS
bound applies even to purely classical systems like a Lennard-Jones
fluid.  \ref{sec:scale} is devoted to the derivation of the
Ward identities that result from scale invariance and the associated
vanishing of the bulk viscosity.  Finally, in
\ref{sec:contact_and_tail_coefficient} we analytically
compute the tail coefficient of the shear viscosity at large
frequencies.

%%%%%%%%%%%%%%%%%%%%%%%%%%%%%%%%%%%%%%%%%%%%%%%%%%%%%%%%%%%%%%%%%%%%%%%%

\section{The unitary Fermi gas and scale invariance}
\label{sec:unitary}

The basic model Hamiltonian that describes a two-component Fermi gas
with contact interactions of strength $\bar g$ is given by
\begin{align}
  \label{eq:1}
  \Hat H = \sum_{\p\sigma} \varepsilon_\p c^\dagger_{\p\sigma}
  c_{\p\sigma} + \bar g(\Lambda)\sum_{\p\p'\q} c_{\p+\q\uparrow}^\dagger
  c_{\p'-\q\downarrow}^\dagger c_{\p'\downarrow} c_{\p\uparrow}\, .
\end{align}
Here $\varepsilon_\p=\p^2/2m$ is the kinetic energy, while $\sigma
=\uparrow, \downarrow$ denotes the two fermion species (hyperfine
levels in the context of cold atoms). Since a contact interaction in
three dimensions needs to be regularized, the coupling strength
$\bar{g}(\Lambda)$ is cutoff-dependent. It is related to the
renormalized physical coupling strength $g=4\pi\hbar^2 a/m$ that is
fixed by the scattering length $a$ via $\bar
g(\Lambda)=g/(1-2a\Lambda/\pi)$.  The unitary gas corresponds to the
special case $g=\infty$ of infinite scattering length.  In order to
see that the system is scale invariant at this point, it is convenient
to model the contact interaction by an exchange of bosons. In a path
integral language, the partition function at temperature $T\equiv
1/\beta$ can then be written as a double integral
\begin{eqnarray}
  \label{eq:Z}
  Z = \int \mathcal{D}(\bar{\psi}_{\sigma},\psi_{\sigma})
  \mathcal{D}(\bar{\psi}_B,\psi_B) e^{-S}
\end{eqnarray}
over Grassmann fields $\bar{\psi}_{\sigma}$, $\psi_{\sigma}$ with
$\sigma=\uparrow,\downarrow$ and a complex scalar field
$\bar{\psi}_B$, $\psi_B$. In the case of a Feshbach resonance, this
field has a direct physical interpretation as a bound fermion pair in
the closed channel.  The action that describes the interacting Fermi
gas in this two-channel description is
\begin{multline}
  \label{eq:twochannel}
  S=\int_0^{\beta}d\tau \int d^3x \bigg[
  \sum_{\sigma={\uparrow,\downarrow}} \bar{\psi}_{\sigma}
  \Big[\hbar\partial_{\tau}-\frac{\hbar^2\nabla^2}{2m}\Big]\psi_{\sigma}
  +\bar{\psi}_{B}\Big[\hbar\partial_{\tau}-\frac{\hbar^2\nabla^2}{4m}+\nu\Big]
  \psi_{B} \\
  + \tilde{g} \Big( \bar{\psi}_{B}\psi_{\uparrow}
  \psi_{\downarrow}+h.c.\Big)\bigg].
\end{multline}

Here, $\nu$ is the detuning of the bosonic field, which allows to
change the induced interaction $g\sim -\tilde{g}^2/\nu$ between the
fermions.  In particular, if $\tilde g$ is held at the fixed point
corresponding to zero effective 
range, $g=\infty$ is reached at resonance
$\nu\equiv 0$, where the bosonic field is massless.  As was shown by
Nikoli\'c and Sachdev \cite{nikolic2007rg}, the action
$S[\psi_\sigma,\psi_B]$ at this point is invariant under scale
transformations $\x\to\lambda\x$ and $\tau\to\lambda^2\tau$.  The
fermionic field $\psi_\sigma(\x,\tau)\to \lambda^{-3/2}
\psi_\sigma(\lambda\x,\lambda^2\tau)$ transforms with the canonical
scaling dimension, while the bosonic field $\psi_B(\vec x,\tau)\to
\lambda^{-2} \psi_B(\lambda\x,\lambda^2\tau)$ acquires an anomalous
dimension.  As a result, the dynamics of the bosons, which arises from
the second term in equation \eqref{eq:twochannel}, is irrelevant at
unitarity.  The description of a finite fermion density requires
adding a contribution $\mu\, \bar\psi_\sigma \psi_\sigma$ to the
action.  With a fixed value of the chemical potential $\mu$, this term
explicitly breaks scale invariance.  For the unitary gas, however, the
relation $\mu\sim\xi(0)k_F^2$ between the chemical potential and the
Fermi wave vector $k_F$ is identical to that of an ideal Fermi gas,
except for a universal constant $\xi(0)\simeq 0.36-0.4$ that describes
the lowering of the Fermi energy due to the attractive interactions
\cite{bloch2008mbp, giorgini2008tuf}.  A scale transformation thus
leaves the action invariant provided $\mu\to \lambda^{-2}\mu$ is
changed accordingly to account for the change in density.
Alternatively, one may introduce the chemical potential through a
special choice for the (common) phase of the fields $\psi_\sigma$ and
$\psi_B$ \cite{son2006general}.  Scale invariance gives rise to a
conserved dilatation current \cite{nishida2007ncf} which implies the
relation $\epsilon=3p/2$ and the vanishing of the bulk viscosity:
during a uniform expansion the system remains self-similar and no
entropy is produced \cite{werner2006unitary, son2007vbv}.  In fact,
the bulk viscosity of the unitary gas vanishes for {\it all}
frequencies \cite{taylor2010viscosity}, a result that will be derived
from an exact Ward identity in \ref{sec:scale}.  A subtle
point in this context is the issue of whether scale invariance of the
action will be present in the full theory, i.e., whether the classical
symmetry of $S$ survives quantum fluctuations
\cite{wess1971consequences, zee2003qft}.  For
the unitary gas with equal masses of the two spin components, which is
the situation discussed in this work, this is believed to be 
the case.  For different masses, 
however, scale invariance will be broken due to the appearance of an 
Efimov effect, i.e., three- and four-body bound states appear even in the absence
of a two-body bound state \cite{petrov2003three, nishida2008universal}.
In particular, four-body bound states appear for mass ratios larger than a
critical value $13.384$ \cite{castin2010four}.  The breaking of scale 
invariance through the appearance of an Efimov effect has been
discussed so far mostly for bosons \cite{braaten2006universality} and
leads to limit cycle flows in a renormalization group treatment
\cite{moroz2009invsquare}.

%%%%%%%%%%%%%%%%%%%%%%%%%%%%%%%%%%%%%%%%%%%%%%%%%%%%%%%%%%%%%%%%%%%%%%%%

\section{Kubo formula for the viscosity}
\label{sec:kubo}  

The unitary Fermi gas is in a fluid state at arbitrary temperatures
and thus has well defined bulk and shear viscosities $\zeta$ and
$\eta$. On a microscopic level, the viscosities are the zero frequency
limits of linear response functions that may be calculated from first
principles by the Kubo formula.  Specifically, both the shear and bulk
viscosities follow from the retarded correlation functions of the
stress tensor $\Pi_{ij}$ (in the zero external momentum limit),
\begin{align}
  \label{eq:kubo}
  \chi^\ret_{ij,kl}(\q=0,\omega)
  = i\int dt\, d^3x\, e^{i\omega t}\, \theta(t)\,
  \bigl\langle [ \Pi_{ij}(\x,t), \Pi_{kl}(\vec 0,0) ] \bigr\rangle \,,
\end{align}
and its imaginary parts $\Im \chi^\ret_{ij,kl}(\omega)$, which are odd
functions of $\omega$. In particular, the real part of the frequency
dependent shear viscosity is determined by the associated positive and
even spectral function
\begin{align}
  \label{eq:etaomega}
  \Re\eta(\omega) & = \frac{\Im \chi^\ret_{xy,xy}(\omega)}{\omega}
\end{align}
and its static limit $\eta = \lim_{\omega\to 0}\Re\eta(\omega)$.  A
completely analogous expression exists for the bulk viscosity
$\zeta(\omega)$ which involves the trace $\Pi_{ii}$ of the stress
tensor,
\begin{align}
  \label{eq:zeta}
  \Re\zeta(\omega) = \frac{\Im\chi_{ii,jj}^\ret(\omega)}{9\omega}\, .
\end{align}
The microscopic expression for the stress tensor operator that enters
equation \eqref{eq:kubo} for a non-relativistic quantum many-body
system has been derived by Martin and Schwinger
\cite{martin1959theory}.  In the case of a scale-invariant short-range
potential, where $-r\partial_rV(r)=2V(r)$, it is given by
\begin{align}
  \label{eq:Tij0}
  \Pi_{ij}(\q=0,t) 
  = \frac {\hbar^2}{m}\, \sum_{\k\sigma} k_i k_j\, c^\dagger_{\k\sigma}
  c^{\phantom\dagger}_{\k\sigma} 
  + \int d^3x \int d^3r\, \frac{r_ir_j}{r^2}\, 2V(\vec r)
  : \! n_\uparrow(\x) n_\downarrow(\x+\vec r) \! : \, .
\end{align}
The second term explicitly involves the interaction and guarantees
that the stress tensor satisfies local momentum current conservation
as an operator equation (see appendix B in Ref.\
\cite{nishida2007ncf}).  In the zero-frequency limit, the kinetic part
of the stress tensor \eqref{eq:Tij0} is the one that appears in the
viscous terms of the hydrodynamic equations \cite{schaefer2009npf},
and also as a source term in the Boltzmann equation
\cite{smith1989transport}.  In general, the two contributions to the
stress tensor \eqref{eq:Tij0} describe physically different processes
for momentum relaxation upon insertion into the Kubo formula
\eqref{eq:kubo}.  In particular, the correlation function of two
interaction contributions describes collisional transport due to
interparticle forces, which is in fact the dominant contribution in
the liquid phase \cite{hansen2006liquids}.  By contrast, the kinetic
part is associated with the transfer of transverse momentum due to
free particle motion and dominates in the gaseous phase.  As will be
shown in section~\ref{sec:calc}, for the unitary gas the interaction
part of the stress tensor is important to guarantee that the bulk
viscosity vanishes, however it gives no contribution to the shear
viscosity due to the zero-range nature of the interaction.
 
Quite generally, momentum current conservation implies that the stress
tensor correlation functions are directly related to the current
correlation functions, which are defined by
\begin{align}
  \label{eq:kubo2}
  \chi^\ret_{i,k}(\q,\omega)
  = i\int dt\, d^3x\, e^{i(\omega t-\q\x)}\, \theta(t)\,
  \bigl\langle [ j_{i}(\x,t), j_{k}(\vec 0,0) ] \bigr\rangle \, ,
\end{align} 
with
\begin{align}
  \label{eq:ji0}
  j_{i}(\q) = 
  \sum_{\k\sigma} \hbar k_i\,
  c^\dagger_{\k-\q/2,\sigma} c^{\phantom\dagger}_{\k+\q/2,\sigma}  
\end{align} 
the current operator. In the limit $\q, \omega\to 0$, these
correlation functions acquire a simple form, dictated by
hydrodynamics.  In particular, the existence of a finite shear
viscosity implies diffusive relaxation of the transverse currents
\cite{hohenberg1965mts, forster1975hydro}.  The transverse part of the
current correlation spectral function thus has the generic form
(for small $\q,\omega$) 
\begin{align}
  \label{eq:diffusion}
  \frac{\Im
    \chi^\ret_\perp(\q,\omega)}{\omega}  
  =\frac{\eta q^2}{\omega^2+\left(D_\eta q^2\right)^2}
\end{align}
with a diffusion constant $D_\eta$ that is directly proportional
to the shear viscosity. Indeed, the sum rule
\begin{align}
  \label{eq:rhon}
  \lim_{\q\to 0}\int\frac{d\omega}{\pi}
  \frac{\Im \chi^\ret_{\perp}(\q,\omega)}{\omega}
  =\rho_n
\end{align}
which quite generally defines the normal fluid density $\rho_n$
\cite{hohenberg1965mts, forster1975hydro} immediately implies an
Einstein relation
\begin{align}
  \label{eq:einstein}
  \eta=\rho_n D_\eta
\end{align}   
between the shear viscosity and the associated diffusion constant that
was first derived by Hohenberg and Martin \cite{hohenberg1965mts}.
The shear viscosity can thus be defined both from the stress tensor
correlation function and, alternatively, from the diffusion constant
that appears in the transverse current response
\cite{forster1975hydro, taylor2010viscosity}.  For non-superfluid
systems, the diffusion constant $D_{\eta}$ is identical to the
standard kinematic viscosity $\nu=\eta/\rho$.

For strongly interacting fluids, an exact calculation of correlation
functions like the ones in equation \eqref{eq:kubo} is hardly
possible.  It is therefore of considerable interest to find
constraints that allow to check the validity of approximate
calculations.  Such constraints are derived generically from a
short-time expansion of the correlation functions that can be
expressed in terms of equilibrium expectation values of certain
commutators.  For the viscosities of a Fermi gas with contact
interactions, this has been achieved recently by Taylor and Randeria
\cite{taylor2010viscosity}. As will be shown in section~\ref{sec:calc}
below, a straightforward derivation of the relevant sum rule can be
given using a Ward identity due to Polyakov \cite{polyakov1969nonequ},
who has discussed the behavior of the shear viscosity near the
critical point of a neutral superfluid within a microscopic approach.
This Ward identity, which follows from momentum conservation, implies
that the frequency-dependent shear viscosity at arbitrary values of
the scattering length $a$ obeys the sum rule
\begin{align}
  \label{eq:sumeta2}
  \frac{2}{\pi} \int_0^\infty d\omega\, \Bigl[ \eta(\omega) -
  \frac{\hbar^{3/2}C}{15\pi\sqrt{m\omega}} \Bigr]
  = \frac{2\varepsilon}{3} - \frac{\hbar^2C}{6\pi ma}
  = p - \frac{\hbar^2C}{4\pi ma} \,.
\end{align}
Here, in the second form of the equality, we have used the non-trivial
relation between energy density and pressure away from unitarity,
first derived by Tan \cite{tan2008large}.  The sum rule
\eqref{eq:sumeta2} has precisely the form given by Taylor and Randeria
\cite{taylor2010viscosity}, however their high-frequency tail coefficient $C_\eta =
C/(10\pi)$ is larger by a factor of $3/2$ and there is also a
difference in the prefactor of the contribution proportional to $C/a$
away from unitarity.  The sum rule implies that $\eta(\omega)$ decays
like $1/\sqrt\omega$ at large frequencies with a prefactor that is
determined by the Tan contact density $C$, which is a measure for two
fermions with opposite spin to be close together
\cite{tan2008energetics, tan2008large, braaten2008exact,
  braaten2008universal}.  Moreover, the area under the
frequency-dependent viscosity of the unitary gas is fixed by the
equilibrium pressure, since the $a^{-1}$ term is absent at unitarity.
Both features are verified with high accuracy by our calculations of
$\eta(\omega)$ that are detailed in section~\ref{sec:results}.

%%%%%%%%%%%%%%%%%%%%%%%%%%%%%%%%%%%%%%%%%%%%%%%%%%%%%%%%%%%%%%%%%%%%%%%%

\section{Temperature dependence of the shear viscosity}
\label{sec:alpha}

For the unitary gas, the density $n=k_F^3/(3\pi^2)$ fixes both the
momentum and energy scale, which are the Fermi wavevector $k_F$ and
the associated Fermi energy $\varepsilon_F=k_BT_F=\hbar^2k_F^2/(2m)$.
Since the viscosity has dimensions $\hbar n$, purely dimensional
arguments require the static shear viscosity to be of the form
\begin{align}
  \label{eq:scaling}
  \eta(T)=\hbar n \alpha(\theta)
\end{align} 
where $\theta=T/T_F$ is the dimensionless temperature scale and
$\alpha(\theta)$ a universal scaling function. Remarkably, this
function is fixed up to a universal constant $\mathcal O(1)$ in the
high temperature limit.  This is based on the counter-intuitive fact
that the viscosity of a classical gas is independent of its density
\cite{balian1992vol2}.  Since $T_F\sim n^{2/3}$, this requires
$\alpha(\theta\gg 1)\sim\theta^{3/2}$, i.e., a shear viscosity that
increases like $T^{3/2}$. The same qualitative result is obtained from
the kinetic theory expression $\eta(T)\simeq\langle
p\rangle/\sigma(T)$ with $\langle p\rangle$ an average momentum and
$\sigma(T)$ a thermally averaged cross section. Since the differential
cross section for a collision with a given relative momentum $p$ is
$d\sigma/d\Omega =\hbar^2/p^2$ at unitarity, this immediately gives
$\eta\simeq p_T^3/\hbar^2$ with $p_T=\sqrt{2\pi mk_BT}$ the thermal
momentum. Note that $\eta$ scales like $1/\hbar^2$ even in the
classical limit, provided the assumption of zero range s-wave
scattering remains valid in this regime.

At low temperatures $\theta\lesssim 0.15$, the unitary Fermi gas is a
superfluid.  Contrary to naive expectations, a superfluid is not a
`perfect fluid' despite the fact that there is a vanishing viscosity
here.  In particular, superfluids do not provide trivial
counter-examples to the KSS conjecture.  Indeed, according to the
Landau two-fluid picture, the superfluid component has both zero
viscosity and zero entropy, so $\eta/s$ is undefined at $T=0$.  At any
finite temperature, however, a normal component appears, whose entropy
and viscosity are non-zero.  For the bosonic superfluid $^4$He this
was discussed already by Landau and Khalatnikov in
1949~\cite{landau1949tvh1}.  In the low temperature, phonon-dominated
regime, they found that the shear viscosity of the normal component
grows like $T^{-5}$ because the mean free path for phonon-phonon
collisions, which are necessary for the relaxation of shear, diverges.
Specifically, with the assumption that the phonon dispersion has
negative curvature and thus no Beliaev decay is possible (this
assumption is actually violated in superfluid $^4$He but is likely to
hold for the unitary Fermi gas, see \cite{diener2008quantum}), the
calculations of Landau and Khalatnikov predict the low temperature
shear viscosity to be
\begin{equation}
  \label{eq:landau}
  \eta(T\to 0)=\rho_n(T)\,\frac{\rho^2
    c_s^3}{\hbar^2}\,\frac{2^{13}(2\pi)^7}{9(13)!(u+1)^4}
  \left(\frac{\hbar\, c_s}{k_BT}\right)^9\, .
\end{equation}
Here $\rho_n(T) =(2\pi^2\hbar/45 c_s)(k_BT/\hbar c_s)^4$ is the normal
fluid mass density and $u=d\ln{c_s}/d\ln{n}$ is the dimensionless
strength of the non-linear corrections to the leading-order quantum
hydrodynamic Hamiltonian which lead to phonon-phonon scattering.  The
viscosity of the normal fluid component thus asymptotically diverges
like $T^{-5}$.  As realized by Rupak and Sch\"afer~\cite{rupak2007svs}
the same behavior is expected in the unitary Fermi gas.  Indeed, at
temperatures far below $T_c$, the microscopic nature of the superfluid
is irrelevant and the linearly dispersing Bogoliubov-Anderson phonons
are the only excitations that remain.  As a result, equation
\eqref{eq:landau} for the viscosity applies also to the unitary Fermi
gas, provided the exact values of the sound velocity $c_s$ and
coupling constant $u$ are inserted.  At unitarity, the sound velocity
$c_s=v_F\sqrt{\xi/3}\simeq 0.36\, v_F$ is directly proportional to the
Fermi velocity $v_F$ with a factor that is determined by the universal
Bertsch parameter $\xi\simeq 0.36-0.4$ \cite{bloch2008mbp}.  Since
$v_F\sim n^{1/3}$ as for an ideal Fermi gas, the dimensionless
coupling constant that fixes the strength of the phonon-phonon
scattering amplitude has the universal value $u=1/3$.  Together with
the standard low-temperature expression $s=2\pi^2k_B/45 \, (k_BT/\hbar
c_s)^3$ for the entropy density of a scalar phonon field, the
viscosity to entropy density ratio
\begin{align}
  \label{eq:landau_ufg}
  \frac{\eta(T\to 0)}{s} = \frac{\hbar}{k_B}\, 2.15 \times
  10^{-5}\, \xi^5 \theta^{-8}
\end{align}
of the unitary Fermi gas at temperatures far below the superfluid
transition diverges rapidly as $(T_F/T)^8$.  The associated prefactor
agrees within $2\%$ with that found from a diagrammatic calculation by
Rupak and Sch\"afer~\cite{rupak2007svs}, which is based on the
next-to-leading terms in the effective field theory of the unitary gas
by Son and Wingate \cite{son2006general} (note the different prefactor
of the entropy in \cite{rupak2007svs}).  It is a pure number that only
contains the Bertsch parameter.  Note that due to $\rho_n(T)/s(T)=T/c_s^2$ at
very low temperatures, the ratio $\eta/sT=D_\eta/c_s^2$ is just the
characteristic relaxation time $\tau_{\eta}$ for shear fluctuations.
According to equation \eqref{eq:landau}, this relaxation time diverges
like $\tau_{\eta}\sim T^{-9}$ as $T\to 0$, much stronger than the pure
thermal energy time scale $\hbar/(k_BT)$.  In fact, the latter would
lead to a viscosity to entropy density ratio which approaches a
constant at very low temperatures, as in the $\mathcal N=4$
supersymmetric Yang-Mills theory, where no quasiparticles exist.

Near the superfluid transition temperature $T_c$, the universal
scaling function $\alpha(\theta)$ will be continuous, yet there will
be singularities in higher derivatives.  Within the conventional
theory of critical dynamics, the non-analytical behavior of
$\eta=\rho_n D_\eta$ near $T_c$ only comes from the thermodynamic
singularity in $\rho_n(T)$ \cite{hohenberg1977dynamic}.  Since
$\rho_n(T)=\rho -c'\,\theta(-t)|t|^{\nu}$ with $c'$ a positive
constant and $\nu\approx 0.67$ the universal critical exponent of the
3D XY-model, the conventional theory predicts that $\alpha(\theta)$
reaches its finite value at $T_c$ with an infinite slope as the
critical temperature is approached from below \cite{ferrell1988ies}.
A non-analytical temperature dependence of the shear viscosity is
indeed observed near the $\lambda$ point of superfluid $^4$He.  In
this case, however, a singularity of the form sign$(t)\,
|t|^\omega$ appears on both sides of the transition
\cite{biskeborn1975critical}.  The associated exponent $\omega$ is
consistent with the prediction $\omega=\nu$ of conventional theory
below $T_c$, however it has a different value $\omega\approx 0.8$
above the critical temperature. This behavior can be explained within
a semi-phenomenological approach \cite{schloms1990csv} which is,
however, genuine for a liquid state.  Thus it cannot be carried over
to the case of the unitary Fermi gas, for which the precise behavior
of $\alpha(\theta)$ near the critical point remains an open problem.

%%%%%%%%%%%%%%%%%%%%%%%%%%%%%%%%%%%%%%%%%%%%%%%%%%%%%%%%%%%%%%%%%%%%%%%%

\section{Diagrammatic evaluation of the stress tensor correlation
  functions}
\label{sec:calc}

\begin{figure}
  \includegraphics[width=.5\linewidth,clip]{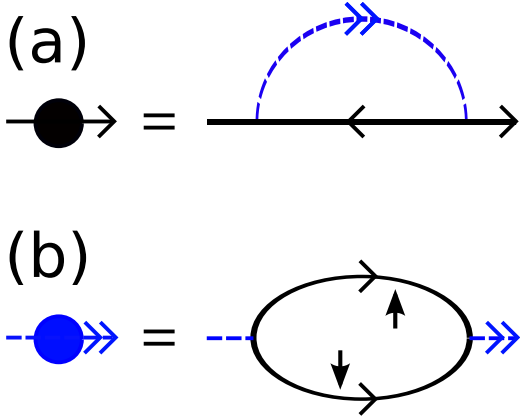}
  \caption{[color online] Diagrammatic contributions to the
    fermionic and bosonic self-energies.}
  \label{fig:sigma}
\end{figure}

In order to compute the stress correlation functions \eqref{eq:kubo}
we start from the single-channel model and derive the corresponding
stress tensor in the two-channel model.  This has the advantage that
the potential (fermion interaction) term of the stress tensor becomes
simply a detuning (mass term) of the bosonic field in the two-channel
model.  We then derive the exact expression \eqref{eq:suscept} for the
stress correlation function in the two-channel model, including both
kinetic and potential terms, in the zero-range limit. 
This expression is evaluated in the
self-consistent T-matrix approximation \cite{baym1961conserv,
  baym1962self}.  In the two-channel model we compute the fermionic
and bosonic self-energies self-consistently at the one-loop level but
neglect the loop corrections to the Yukawa coupling between fermions
and bosons which appear only at two-loop order (cf.\
Fig.~\ref{fig:sigma}).  Variation of these coupled single-particle
equations with respect to a time-dependent external field results in a
set of transport equations for the viscosity response functions.  A
crucial feature of this procedure is that it respects the symmetries
of the underlying model exactly, in particular it obeys scale
invariance and thus leads to a vanishing bulk viscosity as required.

Since the continuum model is Galilean invariant, one may expand the
stress tensor operator \eqref{eq:Tij0} of the single-channel model
\eqref{eq:1} into spherical harmonics with angular-momentum quantum
number $\ell$:
\begin{align}
  \label{eq:Tell}
  \Pi_\ell
  = \sum_{\p\sigma} 2\varepsilon_p Y_\ell(\Hat\p)
  c_{\p\sigma}^\dagger c_{\p\sigma} 
  + 2\delta_{\ell0} \, \bar g(\Lambda) \int d^3r\,
  : \! n_\uparrow(\r) n_\downarrow(\r) \! :
\end{align}
where the spherical harmonics $Y_\ell(\Hat\p)$ depend on the angle of
the vector $\p=(p,\theta,\phi)$ as $Y_{\ell=0}(\theta,\phi)=1$ and
$Y_{\ell=2}(\theta,\phi) = \sin^2(\theta) \sin(\phi) \cos(\phi)$.
(For convenience we omit the standard normalization factor
$1/\sqrt{4\pi}$ in the spherical harmonics.)  For a quadratic
dispersion the fermion kinetic term admits $\ell=0$ and $\ell=2$.
Note that for the scale-invariant, zero-range model at unitarity the
interaction term contributes only for $\ell=0$!

In order to compute linear response one has to add a perturbation to
the Hamiltonian which couples the stress tensor to a time-dependent
external field,
\begin{align}
  \Delta H(t) = \sum_\ell h_\ell(t)\, \Pi_\ell(t) 
  = \sum_{\p\sigma\ell} 2h_\ell(t) \varepsilon_p Y_\ell(\Hat\p)
  c_{\p\sigma}^\dagger c_{\p\sigma} 
  + 2h_{\ell=0}(t)\, \bar g(\Lambda) \int d^3r\,
  : \! n_\uparrow(\r) n_\downarrow(\r) \! : \;.
\end{align}
This amounts to the replacement $\varepsilon_p \mapsto \varepsilon_p
[1+2h_\ell(t)]$ and $\bar g(\Lambda) \mapsto \bar g(\Lambda)
[1+2h_\ell(t)\delta_{\ell0}]$ in the full Hamiltonian, and likewise in
the full action at unitarity
\begin{multline}
  S[h] = \int_0^\beta d\tau\Bigl[ \sum_{\p\sigma}\bigl [1+\sum_\ell
  2h_\ell(\tau) Y_\ell(\Hat\p) \bigr]
  \varepsilon_p \Bar c_{\p\sigma} c_{\p\sigma} + 
  \tilde g \Bar\psi_B \psi_\uparrow \psi_\downarrow + \text{h.c.} \\ 
  - \frac{\tilde g^2}{\Bar g(\Lambda) [1+2h_{\ell=0 }(\tau)]} \,
  \Bar\psi_B \psi_B \Bigr]\,. 
\end{multline}
The change in the action due to the external field, $\Delta S=S[h]-S$,
can be parametrized as
\begin{align}
  \Delta S 
  = \sum_{XX'} \Bigl[ \sum_\sigma \Bar\psi_\sigma(X)
  U_{\sigma,XX'} \psi_\sigma(X') 
  + \Bar\psi_B(X) U_{B,XX'} \psi_B(X') \Bigr] 
\end{align}
with the coefficient functions
\begin{align}
  \label{eq:Usigma}
  & U_{\sigma,XX'} = \int d\tau\, \sum_\ell h_\ell(\tau)
  Y_\ell(\Hat\p) T_{\sigma\ell}^{(0)}(\tau XX') \,,\\
  \label{eq:UB}
  & U_{B,XX'} = \tilde g^2 \int d\tau\, \sum_\ell 
  \frac{h_\ell(\tau)}{1+2h_\ell(\tau)} \,
  S_\ell^{(0)}(\tau XX') \,.
\end{align}
For convenience we use a short-hand notation $X=(\r,\tau)$ for the
real-space argument and $K=(\k,i\epsilon_n)$ for the Fourier argument,
and $\sum_X = \int d^3r\, \hbar^{-1} \int d\tau$ as well as
$\delta_{XX'} = \delta(\r-\r') \hbar \delta(\tau-\tau')$.  Equations
\eqref{eq:Usigma}, \eqref{eq:UB} define the bare fermionic and bosonic
viscosity response vertices at unitarity,
\begin{align}
  \label{eq:T0}
  & T_{\sigma\ell}^{(0)}(\tau X_1X_1') = \frac{\hbar^2\nabla_1\nabla_1'}{m}\,
  \delta_{X_1X_1'} \delta(\tau-\tau_1) \,, \\
  \label{eq:S0}
  & S_\ell^{(0)}(\tau X_1X_1') =
  \begin{cases}
    2 \bar g(\Lambda)^{-1} \delta_{X_1X_1'} \delta(\tau-\tau_1)
    & \text{for $\ell=0$} \\
    0 & \text{for $\ell=2$} \,.
  \end{cases}
\end{align}
The response of the grand potential to the external field in terms of
the fermionic Green's functions $G_{\sigma,XX'} = \vev{\mathcal T
  \psi_\sigma(X) \psi_\sigma^\dagger(X')}$ and bosonic Green's
functions $G_{B,XX'} = \vev{\mathcal T \psi_B(X) \psi_B^\dagger(X')}$
is
\begin{align}
  \delta \Omega
  & = \tr [(\delta U_\sigma) G_\sigma] + \tr [(\delta U_B) G_B] \\
  & = \tr [T_{\sigma\ell}^{(0)} G_\sigma \delta h_\ell(\tau)]
  -\tr [S_\ell^{(0)} \Gamma
  \frac{\delta h_\ell(\tau)}{(1+2h_\ell(\tau))^2}]
\end{align}
where the trace includes the spin sum and in the second line
additionally the $\ell$ sum.  In the second line the bosonic Green's
function is replaced by the vertex function $\Gamma_{XX'} = -\tilde
g^2G_{B,XX'}$.  Hence, we obtain
\begin{align}
  \label{eq:dOmega}
  -\frac{\delta\Omega}{\delta h_\ell(\tau)} 
  = -\sum_{\sigma XX'} T_{\sigma\ell}^{(0)}(\tau XX')\, G_{\sigma,X'X} 
  + \frac{1}{(1+2h_\ell(\tau))^2} \sum_{XX'} S_\ell^{(0)}(\tau XX')\,
  \Gamma_{X'X} \,.
\end{align}
In particular, for a static scaling perturbation
$h_{\ell=0}(\tau)\equiv h$ we recover the Tan energy formula
\cite{tan2008energetics} with the correct UV regularization,
\begin{align}
  \label{eq:tanenergy}
  3p = \vev{\Pi_{ii}}
  = -\left.\frac{d\Omega}{dh} \right\rvert_{h=0} 
  = 2\sum_{\k\sigma} 
  \frac{\hbar^2k^2}{2m} \left( n_{k\sigma} - \frac{C}{k^4} \right) 
  = 2\vev{H} = 2\varepsilon \,,
\end{align}
where the local limit of the vertex function
$\Gamma_{XX'}\rvert_{X'=X^+}$ has been expressed in terms of the
contact density \eqref{eq:Cvtx}. 
The Kubo formula \eqref{eq:kubo} in imaginary time can be re-expressed
in partial-wave components as
\begin{align}
  \chi_\ell(\tau) = \int d^3r\,
  \vev{\mathcal T \Pi_\ell(\r,\tau)\Pi_\ell(\vec 0,0)} \,,
\end{align}
where $\mathcal{T}$ denotes time ordering.  This stress correlation function 
can be obtained from the second derivative of the grand potential \eqref{eq:dOmega},
\begin{align}
  \chi_\ell(\tau)
  & = -\left.\frac{\delta^2\Omega}{\delta h_\ell(\tau) \delta h_\ell(0)}
    \right\rvert_{h_\ell=0} \notag \\
  \label{eq:suscept}
  & = -\tr [T_{\sigma\ell}^{(0)}(0) \Tilde T_{\sigma\ell}(\tau)] 
 + \tr \Bigl[ S_\ell^{(0)}(0) \Bigl( \Tilde S_\ell(\tau)
  -4\delta(\tau)\Gamma(0) \Bigr) \Bigr] 
\end{align}
where the last term in the square brackets comes from the explicit
$h_\ell$ dependence in the second term of equation \eqref{eq:dOmega}.
Note also that this last term is crucial to obtain a vanishing bulk
viscosity at unitarity $a^{-1}=0$, as we will show in
\ref{sec:scale} using the Ward identities that follow from
scale invariance.  $\Tilde T_\ell$ and $\Tilde S_\ell$ are the
fermionic and bosonic viscosity response functions
\begin{align}
  \label{eq:Ttilde}
  & \Tilde T_{\sigma\ell}(\tau XX')
  = \left. \frac{\delta G_{\sigma,XX'}}{\delta h_\ell(\tau)}
  \right\rvert_{h_\ell=0}  
  = \vev{\mathcal T \Pi_\ell(\tau) \psi_\sigma(X)
    \psi_\sigma^\dagger(X')} \,,\\
  \label{eq:Stilde}
  & \Tilde S_\ell(\tau XX')
  = \left. \frac{\delta \Gamma_{XX'}}{\delta h_\ell(\tau)}
  \right\rvert_{h_\ell=0} 
  = -\tilde g^2 \vev{\mathcal T \Pi_\ell(\tau) \psi_B(X)
    \psi_B^\dagger(X')} \,.
\end{align}
Note that for the case of the shear viscosity where $S_\ell^{(0)}=0$,
the viscosity response function can be expressed by the L function
\cite{baym1961conserv} as $\Tilde T_\ell = LT_\ell^{(0)}$, and the
correlation function \eqref{eq:suscept} reduces to the well-known form
$\chi_{\ell=2}(\tau) = -\tr [T_{\sigma,\ell=2}^{(0)}(\tau) L
T_{\sigma,\ell=2}^{(0)}(0)]$.

The formalism developed so far now allows to derive the sum rule
given in equation \eqref{eq:sumeta2}, starting from the Ward identity
for momentum conservation in the static limit of external $\omega=0$,
$\q\to 0$ which reads \cite[eq.\ (5.26)]{polyakov1969nonequ}
\begin{align}
  \label{eq:momcons}
  \Tilde T_{xy} = -p_x \partial G/\partial p_y\, .
\end{align}
Here $\Tilde T_{xy} = \Tilde T_{\ell=2} Y_{\ell=2}$ is the vertex
function of the stress tensor (cf.\ equation \eqref{eq:Ttilde}).
Starting from equation \eqref{eq:suscept} for the shear viscosity
correlation function, we insert the bare shear viscosity vertex
$T_{\ell=2}^{(0)}(p,i\epsilon_n) = \hbar^2p^2/m$ from equation
\eqref{eq:T0} and the full shear viscosity vertex
$T_{\ell=2}^{(0)}(p,i\epsilon_n) = -p \partial G/\partial p$ from the
Ward identity \eqref{eq:momcons} and obtain at zero external Matsubara
frequency $i\omega_m=0$ (note that $\chi_{xy,xy}(i\omega_m) =
\chi_{\ell=2}(i\omega_m)/15$ due to the angular average of
$[Y_{\ell=2}]^2$)
\begin{align*}
  \chi_{xy,xy}(i\omega_m=0)
  & = -\frac{1}{15} \sum_{p\sigma} \frac{1}{\beta} \sum_{i\epsilon_n}
  T_{\ell=2}^{(0)}(p,i\epsilon_n) \Tilde T_{\ell=2}(p,i\epsilon_n) \\
  & = \frac{1}{15} \sum_{p\sigma} \frac{1}{\beta} \sum_{i\epsilon_n}
  \frac{\hbar^2p^2}{m}\, p \, \frac{\partial G(p,i\epsilon_n)}{\partial p} \\
  & = -\frac{2}{15} \sum_{p\sigma} \varepsilon_p \, 
  p \, \frac{\partial n_p}{\partial p}
\end{align*}
in terms of the fermionic momentum distribution $n_p = -\beta^{-1}
\sum_{i\epsilon_n} G(p,i\epsilon_n)$.  We now integrate by parts, with
a boundary term at momentum cutoff $\Lambda$, and employ the Tan
energy formula \cite{tan2008energetics} to express the sum over the
momentum distribution by the internal energy density $\varepsilon$,
\begin{align}
 \chi_{xy,xy}(i\omega_m=0)
  & = \frac{2}{15} \left[ 5\sum_{p\sigma} \varepsilon_p n_p -
    \frac{\hbar^2 p^5 n_p}{2\pi^2 m} \Bigr\rvert_0^\Lambda \right]
  \notag \\
  \label{eq:sumeta}
  & = \frac{2\varepsilon}{3} +
  \frac{4\hbar^2C\Lambda}{15\pi^2m} - \frac{\hbar^2C}{6\pi ma} \,.
\end{align}
If the momentum cutoff $\Lambda$ is translated into a frequency cutoff
$\Omega = \hbar\Lambda^2/m$ \cite[endnote 39]{taylor2010viscosity} we
arrive at the sum rule given in equation \eqref{eq:sumeta2}.

The derivation of the correlation function \eqref{eq:suscept} so far
has been completely general and exact. The remaining challenge then is
to evaluate the viscosity response functions $\Tilde T_\ell$ and
$\Tilde S_\ell$ within the microscopic model for the unitary Fermi
gas.  In the following we shall do this within the T-matrix
approximation.
We start from the Dyson equation for the fermionic Green's function
$G_{\sigma,XX'}$ in the Matsubara formalism,
\begin{align}
  \label{eq:C_Dyson_equation}
  G_{\sigma,XX'}^{-1} = G_{0,\sigma XX'}^{-1} - U_{\sigma,XX'} -
  \Sigma_{\sigma,XX'}
\end{align}
with bare Green's functions $G_0(K)^{-1} =
-(i\hbar\epsilon_n+\mu-\varepsilon_\k)$ and the external field
$U_{\sigma,XX'}$ from equation \eqref{eq:Usigma}.  In the T-matrix
approximation the fermionic self-energy describes how fermions scatter
off pair fluctuations (cf.\ Fig.~\ref{fig:sigma}a),
\begin{align}
  \label{eq:C_self_energy}
  \Sigma_{\sigma,XX'} = G_{-\sigma,X'X} \, \Gamma_{XX'} \,.
\end{align}
The Bethe-Salpeter equation for the vertex function which
mediates the resonant Fermi-Fermi interaction is
\begin{align}
  \label{eq:Gamma}
  \Gamma_{XX'}^{-1} = \bar g(\Lambda)^{-1} - U_{B,XX'}
  + G_{\uparrow,XX'} \, G_{\downarrow,XX'} \,.
\end{align}
It contains the inverse bare coupling $\bar g(\Lambda)^{-1} = g^{-1} -
m\Lambda/(2\pi^2\hbar^2)$, the external field $U_B$ from equation
\eqref{eq:UB}, and the bosonic self-energy (cf.\
Fig.~\ref{fig:sigma}b).  As mentioned above, the dynamics of the pair
field at unitarity $a^{-1}\to 0$ only arises from the excitation of
fermion pairs while the dynamics of the bosons is irrelevant.  Since
we consider a balanced gas with equal populations of fermion species
$\mu_\uparrow=\mu_\downarrow=\mu$ we will henceforth drop the spin
index $\sigma$.

\begin{figure}
  \includegraphics[width=.6\linewidth,clip]{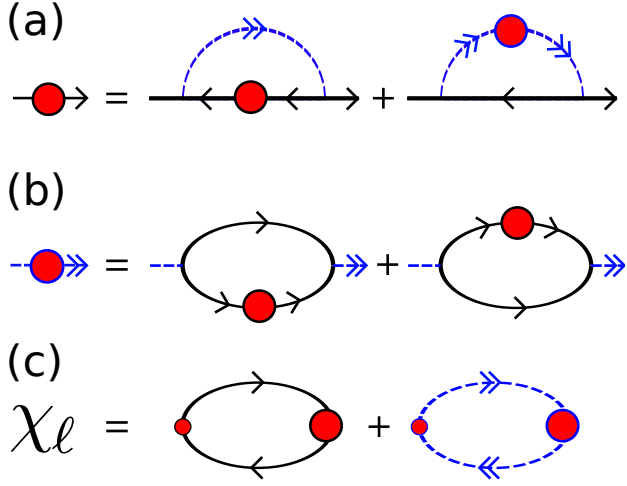}
  \caption{[color online] Diagrammatic contributions to the
    renormalized viscosity response functions $T_\ell$ and $S_\ell$.
    (a) The full fermionic response function $T_\ell$ has
    Maki-Thompson and Aslamazov-Larkin contributions, in addition to
    the bare vertex $T_\ell^{(0)}$.  (b) The full bosonic response
    function $S_\ell$ has two contributions which are equivalent for
    equal fermion masses and populations, in addition to the bare
    vertex $S_\ell^{(0)}$.  (c) The viscosity correlation function is
    given by a skeleton diagram with one bare and one renormalized
    viscosity response vertex and full propagators (fermionic and
    bosonic).  In particular, the second diagram describes transport
    via collective modes.}
  \label{fig:response}
\end{figure}
The T-matrix approximation for the exact viscosity response vertices
is then obtained by taking the derivative of the self-consistency
equations \eqref{eq:C_Dyson_equation}--\eqref{eq:Gamma} with
respect to the external field $h_\ell(\tau)$ (cf.\
Fig.~\ref{fig:response}); in this way it is guaranteed that the
conservation laws are satisfied.  The amputated viscosity response
vertex 
\begin{align}
  \label{eq:C_Fermi_vertex}
  T_\ell(\tau XX') = -\frac{\delta G_{XX'}^{-1}}{\delta h_\ell(\tau)}
  = \sum_{YY'} G_{XY}^{-1}\, \Tilde T_\ell(\tau YY')\, G_{Y'X'}^{-1}
\end{align}
is given by the variation of the Dyson equation
\eqref{eq:C_Dyson_equation} with respect to $h_\ell(\tau)$,
\begin{align}
  \label{eq:C_three_vertex}
  T_\ell(\tau XX') = T_\ell^{(0)}(\tau XX') + T^\MT_\ell(\tau XX') +
  T^\AL_\ell(\tau XX')
\end{align}
with the Maki-Thompson and Aslamazov-Larkin vertex corrections
\begin{align}
  \label{eq:C_vertex_correction_V}
  T^\MT_\ell(\tau XX') & = \frac{\delta G_{X'X}}{\delta h_\ell(\tau)}\,
  \Gamma_{XX'} 
  = \Tilde T_\ell(\tau X'X) \, \Gamma(XX') \,, \\
  \label{eq:C_vertex_correction_AL}
  T^\AL_\ell(\tau XX')
  & = G_{X'X}\, \frac{\delta\Gamma_{XX'}}{\delta h_\ell(\tau)}
  = G_{X'X} \, \Tilde S_\ell(\tau XX') \,.
\end{align}
The amputated bosonic viscosity response vertex 
\begin{align}
  \label{eq:C_Bose_vertex}
  S_\ell(\tau XX') = -\frac{\delta \Gamma_{XX'}^{-1}}{\delta h_\ell(\tau)}
  = \sum_{YY'} \Gamma_{XY}^{-1}\, \Tilde S_\ell(\tau YY')\, \Gamma_{Y'X'}^{-1}
\end{align}
is given by the derivative of the Bethe-Salpeter equation
\eqref{eq:Gamma},
\begin{align}
  \label{eq:visc_bos}
  S_\ell(\tau XX') = S_\ell^{(0)}(\tau XX') -2\, G_{XX'}\, \Tilde
  T_\ell(\tau XX') \,.
\end{align}
The bosonic viscosity vertex describes the response of the pair field
to a scaling or shear perturbation.  The bare bosonic viscosity vertex
$S_\ell^{(0)}$ serves to remove the UV divergence of the
particle-particle loop (second term) which is present only for the
bulk viscosity $\ell=0$.

The self-consistent transport equations for the two-channel model are
equivalent to those of the standard T-matrix approximation in a purely
fermionic description.  Specifically, the two fermionic vertex
corrections $T_\ell^\MT$ and $T_\ell^\AL$ above are precisely those
arising in the self-consistent solution of the integral kernel of the
$L$ function, as introduced by Baym \cite[equation
(60)]{baym1961conserv}.  In physical terms, the first term describes
the interaction between two fermions by exchange of a single pair
while in the second term two pairs are exchanged.  In the context of
calculating the change in conductivity due to superconducting
fluctuations, these terms are called the Maki-Thompson (MT) and
Aslamazov-Larkin (AL) contributions respectively \cite{aslamazov1968},
a notation that will be used also in the present context.  Note that
the second term explicitly includes particle-hole fluctuations, and it
is also referred to as the ``box diagram'' in a functional
renormalization group approach to the thermodynamics of the unitary
Fermi gas \cite{floerchinger2008particle}.  These vertex corrections
are in fact crucial to obtain an approximation which satisfies the
conservation laws of the underlying model.

The self-consistent equations
\eqref{eq:C_Dyson_equation}--\eqref{eq:visc_bos} have a structure
similar to the equations of the Luttinger-Ward approach to the
BCS-BEC crossover developed in our previous work
\cite{haussmann2007tbb, haussmann2009spectral}.  In particular, using
Fourier transforms and the convolution theorem they become algebraic
equations which afford an efficient numerical solution.  
The numerical calculations are performed in three steps.  In the first
step the Green's function $G_{XX'}$ and the vertex function
$\Gamma_{XX'}$ are calculated by solving the self-consistent equations
\eqref{eq:C_Dyson_equation}--\eqref{eq:Gamma} iteratively.  Without
the external fields $U_\sigma$ and $U_B$ the Dyson and Bethe-Salpeter
equations are diagonal in Fourier space, while the fermionic and
bosonic self-energies are local in real space.  Hence, the coupled
equations are solved efficiently by going back and forth between real
and Fourier space.

In the second step $G_{XX'}$ and $\Gamma_{XX'}$ are used as input for
the self-consistent equations
\eqref{eq:C_Fermi_vertex}--\eqref{eq:visc_bos} to calculate the
viscosity response functions $\Tilde T_\ell$, $\Tilde S_\ell$.  Again,
the integral equations \eqref{eq:C_Fermi_vertex} and
\eqref{eq:C_Bose_vertex} become algebraic and are solved in Fourier
space, while the other equations remain local in real space.  Note
that the spatial Fourier transform between radial distances $r$ and
radial wavenumber $k$ for the partial-wave component $\ell$ is given
by
\begin{align}
  T_\ell(k) &= 4\pi (-i)^\ell \int_0^\infty dr\, r^2\, j_\ell(kr)\,
  T_\ell(r) \,, \\
  T_\ell(r) &= \frac{i^\ell}{2\pi^2} \int_0^\infty dk\, k^2\, j_\ell(kr)\,
  T_\ell(k) \,.
\end{align}
In the third step the correlation function $\chi_\ell(i\omega_m)$ is
computed from \eqref{eq:suscept}.  It is continued analytically from
the discrete imaginary Matsubara frequencies $i\omega_m$ to the
continuous real frequencies $\omega$ via both the Pad\'e method and a
model fit function (cf.\ section \ref{sec:results}).  We thus obtain
the retarded correlation function $\chi_\ell^\ret(\omega) =
\chi_\ell^\prime(\omega) + i\, \chi_\ell^{\prime\prime}(\omega)$.
Finally, the real parts of the viscosities $\eta$ and $\zeta$ are
obtained from the correlation functions for $\ell=2$ and $\ell=0$
according to (cf.\ equations \eqref{eq:etaomega} and \eqref{eq:zeta})
\begin{align}
  \label{eq:etasusc}
  \Re\eta(\omega) & = \frac{\Im\chi_{\ell=2}^\ret(\omega)}{15\omega} \,,\\
  \label{eq:zetasusc}
  \Re\zeta(\omega) & = \frac{\Im\chi_{\ell=0}^\ret(\omega)}{9\omega} \,,
\end{align}
where the prefactor of $\eta$ comes from the angular integration of
the spherical harmonics $[Y_{\ell=2}(\Hat\p)]^2$.  Alternatively, one
may solve the integral equation directly for real frequencies where
the limit $\omega\to0$ can be taken analytically.  In practice, this
approach is useful at high temperatures, where self-consistency no
longer plays a role.

%%%%%%%%%%%%%%%%%%%%%%%%%%%%%%%%%%%%%%%%%%%%%%%%%%%%%%%%%%%%%%%%%%%%%%%%

\section{Boltzmann-equation limit}
\label{sec:boltzmann}

In the high-temperature limit $T\gg T_F$ the integral equations
\eqref{eq:C_Fermi_vertex}--\eqref{eq:visc_bos} can be solved by
expanding in powers of the fugacity
\begin{align}
  z = e^{\beta\mu}
  = \frac{4}{3\sqrt\pi} \theta^{-3/2} + \mathcal O(\theta^{-3}) \,.
\end{align}
To leading order in $z$, the pair propagator and on-shell self-energy
are given by
\begin{align}
  \label{eq:Gammaret}
  & \Gamma^\ret(\k,\Omega) = -i\, \frac{4\pi\hbar^3m^{-3/2}}
  {\sqrt{\hbar\Omega+2\mu-\varepsilon_\k/2}} + \mathcal O(z) \\
  \label{eq:sigmaret}
  & \Sigma^\ret(\p,\epsilon=\varepsilon_\p-\mu)
  = i\, \frac{8\varepsilon_F}{3\pi}\,
  \frac{\erf(\sqrt\pi p/p_T)}{p/p_F} + \mathcal O(z) \,.
\end{align}
In the case of on-shell fermions with $\k=\p_1+\p_2$,
$\hbar\Omega+2\mu = \varepsilon_{\p_1}+\varepsilon_{\p_2}$ the pair
propagator reduces to the well-known scattering amplitude $f(q)=i/q$
at infinite scattering length of two particles in vacuum, with
relative momentum $q$.  Note that the exact leading-order result for
the on-shell fermionic self-energy contains a non-trivial
error-function dependence on the ratio of the momentum $p$ to its
thermal value $p_T$ that was missing in previous studies
\cite{combescot2006self}.  It is due to the square-root tail in the
pair propagator and gives a noticeable correction at thermal momenta
$p \simeq p_T$.  Moreover, this form is indeed crucial to fulfill the
condition of scale invariance, as will be discussed below.

The fermionic spectral function in the low fugacity, high temperature
limit has most of the spectral weight concentrated in the coherent
peak at $\epsilon = \varepsilon_\p-\mu$.  The peak width $\gamma_p
=\Im \Sigma^\ret(\p,\epsilon)$ vanishes like $\varepsilon_F\,
p_F/p\sim T^{-1/2}$ for typical momenta $p\approx p_T$, consistent
with the assumption for the temperature dependence of the relaxation
time introduced by Bruun and Smith \cite{bruun2007sva}.  This implies,
in particular, that the fermionic quasiparticles become well-defined
and thus a Boltzmann equation description is valid in the regime
$\theta\gg 1$.

From a numerical, iterative solution of the integral equations
\eqref{eq:C_Fermi_vertex}--\eqref{eq:visc_bos} in the high-temperature
limit we obtain $\eta/(\hbar n) = 2.80(1)\, (T/T_F)^{3/2}$.  This
fixes the constant in the asymptotic behavior
$\alpha(\theta)=\textit{const}\; \theta^{3/2}$ at large values of
$\theta$ of the universal function introduced in \eqref{eq:scaling}.
Within the error bars, the numerical value agrees with that obtained
from a variational solution of the full Boltzmann equation, using
higher Sonine polynomials \cite[appendix]{bruun2007sva}.  The
prediction of a simple power-law dependence of the shear viscosity
$\eta(T)\sim T^{3/2}$ has recently been verified experimentally in a
temperature range between $\theta\simeq 1.5$ and $\theta\simeq 7$ by
measuring the expansion dynamics of a unitary gas released from an
optical trap \cite{cao2010observation}.  Very good agreement has been
found also with the expected prefactor, thus considerably improving
the situation compared to earlier measurements of the shear viscosity from 
the damping of the radial breathing mode \cite{turlapov2008gas}.

\begin{figure}
  \includegraphics[width=\linewidth,clip]{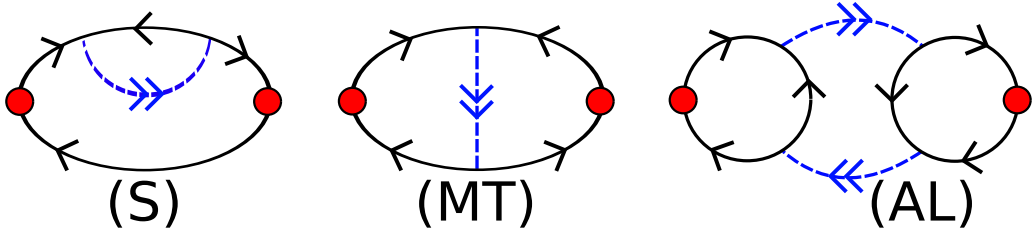}
  \caption{[color online] Diagrammatic contributions to the viscosity
    correlation function $\chi_\ell(\omega)$ at first order in the
    pair fluctuations: Self-energy (S), Maki-Thompson (MT) and
    Aslamazov-Larkin (AL) diagrams.}
  \label{fig:firstorder}
\end{figure}

Remarkably, the solution of the transport integral equation at high
temperatures and small frequencies can also be obtained by a
completely analytical approach.  In fact, in the low fugacity limit,
one can terminate the iterative procedure after the first iteration
step (correlation function to first order in the pair propagator) and
resum via a memory function approach, a method that was developed in
the context of electrical conductivities by G\"otze and W\"olfle
\cite{goetze1972}.  The first-order correlation function contains the
diagrams for self-energy, Maki-Thompson and Aslamazov-Larkin
contributions shown in Fig.~\ref{fig:firstorder}.  These diagrams are
obtained by evaluating the transport equations
\eqref{eq:C_Fermi_vertex}--\eqref{eq:visc_bos} with the bare viscosity
vertices $T_\ell^{(0)}$, $S_\ell^{(0)}$ on the right-hand side, and
with bare fermionic propagators $G_0(p,\epsilon)$ with an additional
impurity scattering rate $\gamma$, which is taken to zero after
resummation.  

Explicitly, to lowest order both in the fugacity and in the scattering
rate, the bosonic vertex correction $S_\ell$ in equation
\eqref{eq:visc_bos} is given by
\begin{align*}
  & S_{\ell=2}(k,\Omega+i0,\Omega-i0)
  = \frac{\varepsilon_F^{1/2}\varepsilon_k
    \sqrt{2(\hbar\Omega+2\mu)-\varepsilon_k}}
  {32\pi\gamma} \,. 
\end{align*}
Here, $S_\ell$ in Fourier space (for notation see equation
\eqref{eq:WIferm}) is continued to real frequency, with one retarded
and one advanced boson line, $S_\ell(k,\omega+i0, \Omega+\omega+i0,
\Omega-i0) \to S_\ell(k,\Omega+i0,\Omega-i0)$ in the limit of
vanishing external frequency $\omega\to0$, to obtain a function of
wavenumber $k$ and real frequency $\Omega$.  This bosonic viscosity
vertex is used to compute the fermionic AL vertex correction in
equation \eqref{eq:C_vertex_correction_AL}.

Next, the fermionic MT and AL viscosity vertices are continued to
real frequencies, with one retarded and one advanced fermion line,
$T_\ell(k,\omega+i0, \epsilon+\omega+i0, \epsilon-i0) \to
T_\ell(k,\epsilon+i0,\epsilon-i0)$ in the limit of vanishing external
frequency $\omega\to0$.  Finally, since the fermions are well-defined
quasiparticles it is sufficient to evaluate the viscosity vertices
on-shell ($\epsilon=\varepsilon_\p-\mu$).  Introducing a dimensionless
momentum variable $y=p/\sqrt{2mk_BT}=\sqrt\pi\, p/p_T$, the resulting
fermionic MT and AL vertex corrections are given by
\begin{align*}
  & T_{\ell=2}^\MT(y)
  = \frac{zT^2}{\sqrt\pi\gamma} \Bigl[ -\frac{3}{4y^3}\erf(y) +
    \frac{1}{\sqrt\pi} \bigl( 1+\frac{3}{2y^2} \bigr) e^{-y^2} \Bigr]
  \\
  & T_{\ell=2}^\AL(y)
  = \frac{zT^2}{\sqrt\pi\gamma} \Bigl[ 
  \bigl(y+\frac 1y+\frac{3}{4y^3}\bigr) \erf(y) 
  -\frac{3}{\sqrt\pi} \bigl( 1+\frac{3}{2y^2}
  \bigr) e^{-y^2} \Bigr].
\end{align*}
From the two fermionic vertex corrections and the self-energy
contribution one obtains the leading correction to the shear viscosity
$\eta(\omega\to 0)$ according to Fig.~\ref{fig:firstorder},
\begin{align}
  \eta^{(1)} & = \eta_\SE^{(1)} + \eta_\MT^{(1)} + \eta_\AL^{(1)}
  = \frac{z^2(2m)^{3/2}T^{7/2}}{15\pi^{5/2}\hbar^2\gamma^2} \times
  \frac{-43-3+30}{8\sqrt 2}
  = -\frac{\sqrt 2z^2(2m)^{3/2}T^{7/2}}{15\pi^{5/2}\hbar^2\gamma^2}
\end{align}
which is negative because adding a scattering channel on top of the
impurity scattering $\gamma$ lowers the viscosity.  Now, the memory
function approach effectively assumes that the first-order correction
$\eta^{(1)}$ to the non-interacting viscosity
\begin{align}
  \eta^{(0)} & = \frac{z(2m)^{3/2}T^{5/2}}{8\pi^{3/2}\hbar^2\gamma}
\end{align}
is the first term of a geometric series, 
\begin{align}
  \eta & = \eta^{(0)} + \eta^{(1)} + \dotsm 
  \approx \eta^{(0)} \left[ 1-\eta^{(1)}/\eta^{(0)} \right]^{-1}
  \notag \\
  & \overset{\gamma\to 0}{\longrightarrow} -\frac{\left(
      \eta^{(0)}\right)^2}{ \eta^{(1)}}
  = \frac{15\hbar k_F^3}{64\sqrt{2\pi}} \, \theta^{3/2}\, .
\end{align}
Normalized to the particle density, the resulting shear viscosity at
high temperatures is thus given by
\begin{align}
  \label{eq:etaclass}
  \eta = \frac{45\pi^{3/2}}{64\sqrt 2}\, \hbar n\, \theta^{3/2}
  \approx 2.77\, \hbar n\, \theta^{3/2} \,.
\end{align}
This analytical result is in perfect agreement with that obtained from
the numerical solution of the full transport equations above and
agrees also with the result obtained from the Boltzmann equation,
using the standard variation of the distribution function $\delta f
\propto v_xp_y$ \cite{massignan2005vra, bruun2005vat}.  Our evaluation
of the Kubo formula within the T-matrix approximation thus recovers
the exact Boltzmann equation result at high temperatures and low
frequencies.  It is important to note that a calculation of the shear
viscosity which uses a nontrivial Ansatz for the fermionic spectral
functions but does not include vertex corrections \cite{bruun2007sva},
gives the correct $T^{3/2}$ asymptotic power law at unitarity, however
the associated prefactor $1.06$ is far too small.  Indeed, the
single-particle scattering time underestimates the transport
scattering time $\tau_\eta$ by a factor of almost $2.6$.  The V and AL
vertex corrections in the Kubo formula are therefore crucial even in
the classical limit.  They are equivalent to solving the full
Boltzmann equation, not only its relaxation-time approximation.

Moreover, neglecting vertex corrections violates the Ward identities
associated with scale invariance and thus leads to a finite bulk
viscosity at unitarity.  Indeed, it is straightforward to show that
only the sum of the three first-order corrections to the bulk
viscosity (order $\gamma^{-2}$ in the impurity scattering rate)
\begin{align}
  \zeta^{(1)} & = \zeta_\SE^{(1)} + \zeta_\MT^{(1)} + \zeta_\AL^{(1)}
  \equiv 0
\end{align}
vanishes identically, as expected for a scale-invariant system.  The
error-function term in the fermionic self-energy \eqref{eq:sigmaret}
is crucial for this cancellation, as well as for satisfying the Ward
identities derived in \ref{sec:scale}.

%%%%%%%%%%%%%%%%%%%%%%%%%%%%%%%%%%%%%%%%%%%%%%%%%%%%%%%%%%%%%%%%%%%%%%%%

\section{Results for the shear viscosity}
\label{sec:results}

As the temperature is lowered towards the degeneracy temperature
$T_F$, the fugacity ceases to be a good expansion parameter, reaching
$z=1$ at around $T\approx 0.6\, T_F$.  In a perturbative approach,
this regime can be approached from the ideal gas via the virial
expansion \cite{ho2004high, werner2006threebody, liu2009virial}.
The known exact values of the second and third virial coefficient can
then be used for a calibration of measurements of the equation of
state of the unitary gas \cite{luo2009thermodynamic,
  nascimbene2010exploring, horikoshi2010measurement}. 
For a calculation of dynamical 
properties like the viscosity in the regime $0.15<\theta\lesssim 1$,
where the gas is degenerate but still in a non-superfluid state, it is
necessary to use a non-perturbative approach that also provides
information about dynamical correlation functions.  As discussed in
section~\ref{sec:calc}, this can be achieved by a self-consistent
Luttinger-Ward formulation of the many-body problem, which allows to
determine not only thermodynamic properties of the unitary gas at
arbitrary temperatures \cite{haussmann2007tbb} but also the fermionic
spectral functions \cite{haussmann2009spectral}.  In the relevant
regime $0.15<\theta\lesssim 1$ just above the superfluid transition
temperature, they exhibit a substantial broadening, which is of
order of the Fermi energy itself \cite{haussmann2009spectral}.  As a
result, the fermionic quasiparticles are no longer well-defined.  The
integral equations \eqref{eq:C_Fermi_vertex}--\eqref{eq:visc_bos} for
the viscosity vertices as functions of (radial) distance and imaginary
time can then only be solved numerically.  In practice, the
self-consistent set of integral equations is iterated until
convergence is reached (typically after a few steps); then the
correlation function
$\chi_{xy,xy}(i\omega_m)=\chi_{\ell=2}(i\omega_m)/15$ is continued
analytically to real frequencies.

\begin{figure}
  \includegraphics[angle=-90,width=.8\linewidth,clip]{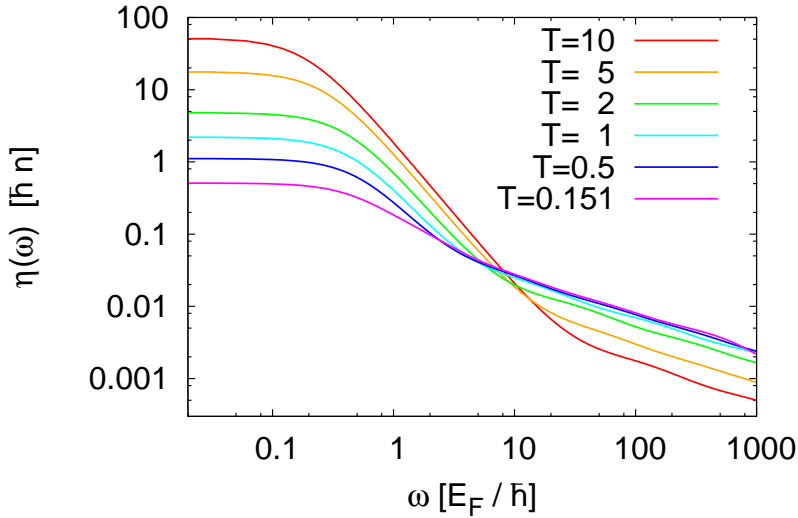}
  \caption{[color online] Shear viscosity spectral function
    $\eta(\omega)$ for different temperatures (on the left side of the
    plot, temperature decreases from top to bottom; on the right side
    the order is opposite).  The analytical continuation is performed
    by the Pad\'e method.}
  \label{fig:etapeak}
\end{figure}

As a first step to evaluate the shear viscosity for real frequencies,
we perform the analytic continuation by the Pad\'e method, using the
first 300 Matsubara frequencies \cite{haussmann2009spectral}.  The
resulting $\eta(\omega)$ curves are shown in Fig.~\ref{fig:etapeak}.
They exhibit a clear Drude peak, which crosses over into a
$\omega^{-1/2}$ tail at large $\omega$.  Unfortunately, it is
difficult to extract a precise value for the transport scattering time
at large temperatures by this method.  To obtain reliable results for
the frequency-dependent viscosity, we therefore use a simple model
function of the form
\begin{align}
  \label{eq:modelfct}
  \chi_{xy,xy}(i\omega_m) 
  = \frac{W-\sqrt 2\, \hbar^{3/2} C_\eta m^{-1/2} \tau_\eta
    \abs{\omega_m}^{3/2}} 
  {1+\tau_\eta \abs{\omega_m}} + \frac{4\hbar^2C\Lambda}{15\pi^2m} \, .
\end{align}
It describes the dependence on the Matsubara frequencies found
numerically from the self-consistent solution of the transport
equation very well for arbitrary temperatures above $T_c$.  The
additive contribution in equation \eqref{eq:modelfct} which is linear
in the momentum cutoff $\sim C\Lambda$ has been derived in
\eqref{eq:sumeta}.  It is, however, irrelevant for the associated
viscosity spectral function
\begin{multline}
  \eta(\omega)
  = \frac{\Im \chi^\ret_{xy,xy}(\omega)}{\omega}
  =\frac{W\tau_\eta}{1+(\omega\tau_\eta)^2} 
  + \frac{\hbar^{3/2}C_\eta}{\sqrt{m\omega}} 
  \, \frac{\omega\tau_\eta(1+\omega\tau_\eta)} {1+(\omega\tau_\eta)^2} 
  \label{eq:modellimit} \\
  \longrightarrow
  \begin{cases}
    W\tau_\eta + \hbar^{3/2} C_\eta \tau_\eta \sqrt{\omega/m} 
    & \text{ as } \omega\to 0 \\
    \hbar^{3/2}C_\eta/\sqrt{m\omega} 
    & \text{ as } \omega\to\infty
  \end{cases} 
\end{multline}
which exhibits a Drude-like transport peak and a $1/\sqrt{\omega}$
tail, as expected on general grounds.  This viscosity function is in
very good agreement with the Pad\'e results shown in
Fig.~\ref{fig:etapeak} and has the advantage that the resulting
$\tau_\eta$ from the fitting function agrees perfectly with the exact
result in the classical limit.  Within this Ansatz, the 
frequency-dependent shear viscosity is characterized by three
parameters: the total Drude weight $W(T)$, the viscous transport
scattering time $\tau_\eta(T)$ and a tail coefficient $C_\eta(T)$.
In practice, since both the Drude weight and the tail coefficient
are fixed by the equilibrium variables $p$ and $C$ via the exact sum
rule \eqref{eq:sumeta2}, 
the transport scattering time remains as the single adjustable parameter. 
Note that the dc-viscosity in equation \eqref{eq:modellimit} has
precisely the form that is assumed in Maxwell's theory of strongly
viscous fluids, with $W\equiv G_{\infty}$ playing the role of the
high-frequency shear modulus \cite{hansen2006liquids}.  An unexpected
feature of our model function \eqref{eq:modellimit} is the presence of a
$\sqrt{\omega}$ singularity at low frequencies, whose weight is too small,
however, to be seen in Fig.~\ref{fig:etapeak}.  As a result, there is a 
negative $t^{-3/2}$ long-time tail in the relaxation of shear stress, similar
to that arising in both classical and quantum liquids due to mode coupling effects
\cite{hansen2006liquids, kirkpatrick2002long}.  If we
insert the model function into the left-hand side of the sum rule 
\eqref{eq:sumeta2},
\begin{align*}
  \frac{2}{\pi} \int_0^\infty d\omega\, \Bigl[
  \frac{W\tau_\eta}{1+(\omega\tau_\eta)^2}  
  + \frac{\hbar^{3/2}C_\eta}{\sqrt{m\omega}} 
  \, \frac{\omega\tau_\eta(1+\omega\tau_\eta)} {1+(\omega\tau_\eta)^2} -
  \frac{\hbar^{3/2}C}{15\pi\sqrt{m\omega}} \Bigr]
  = W ,
\end{align*}
the second term of the model function and the subtraction of the
high-frequency tail integrate to zero exactly if $C_\eta = C/(15\pi)$,
for any value of $\tau_\eta$.  The frequency integral thus yields the
Drude weight $W$, which is equal to the equilibrium pressure $p$ at
unitarity by the exact sum rule \eqref{eq:sumeta2}.

\begin{figure}
  \includegraphics[angle=-90,width=.8\linewidth,clip]{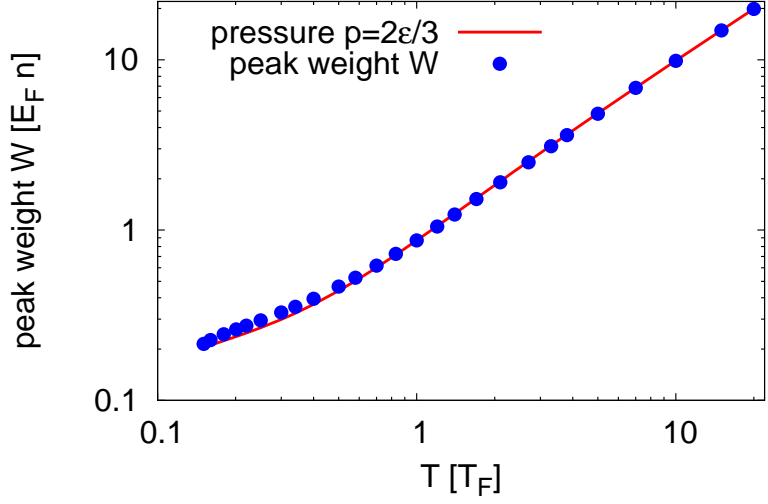}
  \caption{[color online] Weight $W$ of the Drude peak in the shear
    viscosity spectral function $\eta(\omega) \approx
    W\tau_\eta/[1+(\omega\tau_\eta)^2]$ as a function of temperature.
    The pressure $p(T)$ (red line) is from
    Ref.~\cite{haussmann2007tbb}.}
  \label{fig:W}
\end{figure}
As shown in Fig.~\ref{fig:W}, the total weight of the Drude peak $W$
agrees remarkably well with the equilibrium pressure of the
unitary gas for 
all temperatures, from the classical limit down to the superfluid
transition at $T_c\simeq 0.15\, T_F$.  Here and in the following, all
thermodynamic properties of the unitary Fermi gas are taken from the
equivalent Luttinger-Ward calculation of our previous work in Ref.\
\cite{haussmann2007tbb}.
\begin{figure}
  \includegraphics[angle=-90,width=.8\linewidth,clip]{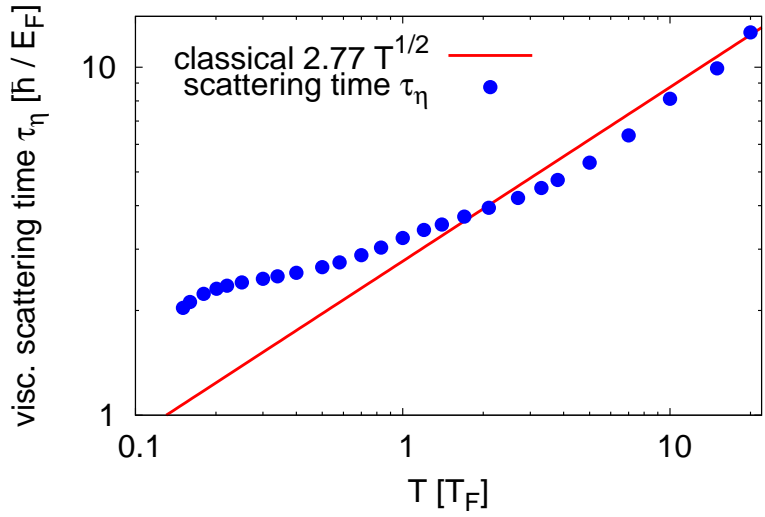}
  \caption{[color online] Viscous scattering time $\tau_\eta$ of the
    Drude peak in the shear viscosity spectral function $\eta(\omega)
    \approx W\tau_\eta/[1+(\omega\tau_\eta)^2]$ as a function of
    temperature.  The classical limit (red line) is from equation
    \eqref{eq:etaclass}.}
  \label{fig:tau}
\end{figure}
The viscous transport scattering time $\tau_\eta$ versus temperature
is shown in Fig.~\ref{fig:tau}: for large temperatures it approaches
$\tau_\eta \sim \theta^{1/2}\,\hbar/\varepsilon_F$, consistent with
equation \eqref{eq:etaclass} and the result obtained from kinetic
theory \cite{bruun2007sva}.  For temperatures near $T_c$, it is
enhanced by a factor of two compared with a simple extrapolation of
the high temperature result.  Qualitatively, there are two competing
effects on the scattering time as the temperature is lowered from the
classical limit: Pauli blocking reduces the phase space and increases
$\tau_\eta$, while pairing fluctuations enhance scattering and lower
$\tau_\eta$.  In a kinetic theory approximation, both effects nearly
balance and $\tau_\eta$ remains essentially at the classical value
down to $T_c$ \cite{bruun2009feshbach}.  By contrast, as is evident
from Fig.~\ref{fig:tau}, our evaluation of the Kubo formula within a
self-consistent T-matrix approximation predicts strong deviations from
kinetic theory in the relevant regime $\theta\lesssim 1$.  Kinetic
theory is clearly inapplicable for a degenerate gas and indeed the
width $\hbar/\tau_\eta \approx 0.5\, k_BT_F$ is more than a factor
three larger than the thermal energy at the lowest temperature $T=T_c$
that is studied here.

\begin{figure}
  \includegraphics[angle=-90,width=.8\linewidth,clip]{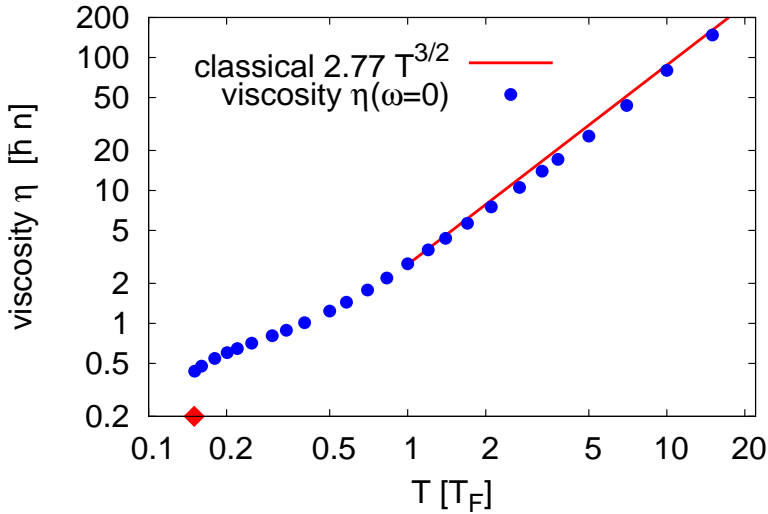}
  \caption{[color online] Static shear viscosity $\eta(\omega=0)$ as a
    function of temperature.  The classical limit $\eta \simeq 2.77\,
    (T/T_F)^{3/2}$ (red line) is from equation \eqref{eq:etaclass}.
    The red diamond indicates $T_c \simeq 0.15\,T_F$.}
  \label{fig:eta}
\end{figure}
The static value $\eta=\eta(\omega=0)$ of the shear viscosity is shown
in Fig.~\ref{fig:eta}.  It exhibits a monotonic dependence on
temperature, reaching $\eta\sim 0.5\, \hbar n$ at the superfluid
transition temperature, still with a positive slope.  As discussed in
section~\ref{sec:alpha}, one expects a diverging positive derivative
of $\eta(T)$ just below $T_c$.  As a result, the minimum of the
viscosity is expected below the superfluid transition, a regime that
is not accessible within our present approach.  The strong rise of the
shear viscosity predicted for $T\ll T_c$ in equation
\eqref{eq:landau}, however indicates that the minimum value of the
shear viscosity for the unitary gas is close to the value reached at
$T_c$.  The dimensionless function $\alpha(\theta)$ thus has a minimum
value of order $\alpha_\eta^{\text{min}} \simeq 0.5$, a value that is
smaller than any other that is found for non-relativistic
fluids~\cite{schaefer2009npf}.  Recalling the connection
\eqref{eq:einstein} between shear viscosity and the associated
diffusion constant $D_\eta$, which is identical with the kinematic
shear viscosity $\nu$ for a normal fluid, the minimum value for $\eta$
is equivalent to a minimum value $D_\eta\simeq 0.5\, \hbar/m$ of the
shear diffusion constant, which only involves Planck's constant and
the particle mass.  Bounds of a similar form are also expected for
other diffusion constants in the unitary gas, for instance heat or
particle diffusion.  In fact, the latter has been measured recently by
studying the equilibration dynamics after the two spin components are
separated in a trap \cite{zwierlein2010diffusion}.  In the degenerate
regime, the observed diffusion constant is again of order $\hbar/m$.
Concerning heat transport, which so far has neither been calculated
nor measured for the unitary gas, it is interesting to note that the
associated diffusion constant $D_T$ is predicted to be identical with
the shear diffusion constant $D_\eta$ for non-relativistic fluids that
exhibit a gravity dual \cite{rangamani2009conformal}.  A comparison of
thermal diffusion and shear viscosity would thus provide a measure of
how close the unitary gas is to such an idealized system, whose
Prandtl number $\text{Pr}=D_\eta/D_T$ is identically equal to one at
all temperatures. In the non-degenerate regime $\theta\gg 1$ one
expects in fact that $\text{Pr}\approx 2/3$, a result that is
essentially universal for dilute classical gases
\cite{smith1989transport}.

\begin{figure}
  \includegraphics[angle=-90,width=.8\linewidth,clip]{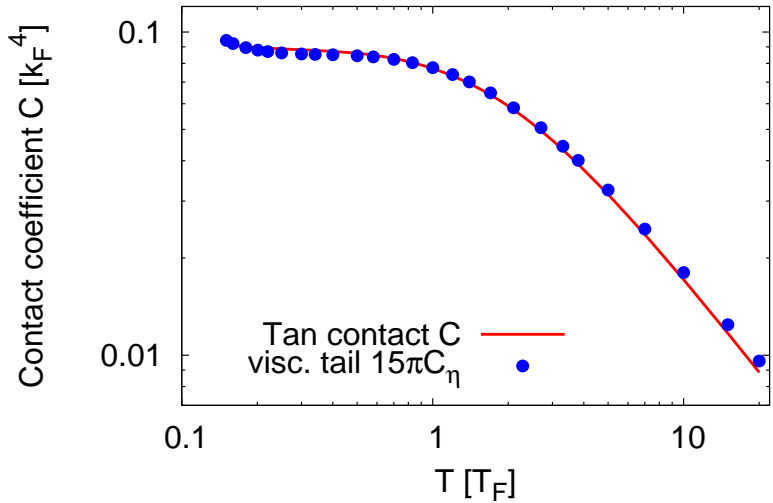}
  \caption{[color online] Tail of the shear viscosity spectral
    function $\eta(\omega) \approx \hbar^{3/2}C_\eta/\sqrt{m\omega}$
    as a function of temperature.  The contact $C$ (red line) is
    computed numerically according to
    Ref.~\cite{haussmann2009spectral} and follows the high-temperature
    asymptotics in equation \eqref{eq:Chightemp}.}
  \label{fig:etatail}
\end{figure}
The tail in the viscosity spectral function \eqref{eq:modellimit}
has the general form 
\begin{align}
  \label{eq:etatail}
  \eta(\omega) & = \frac{\hbar^{3/2}C_\eta}{\sqrt{m\omega}}
  + \mathcal O(\omega^{-3/2})
  && (\omega \gg \omega_x)
\end{align}
above a crossover scale $\omega_x$ which we define as the intersection
point of the Drude and square-root asymptotics.  The crossover
frequency $\omega_x(T)$ at large temperatures scales like $\omega_x
\sim T$ which agrees with the extension of the Drude peak expected
from kinetic theory, $\omega \leq T$ \cite{schaefer2009npf}.  Note
that near $T_c$ the peak viscosity is still about five times higher
than the viscosity $\eta(\omega_x)$ at the onset of the tail, but the
transport peak is nevertheless much less pronounced as compared to higher
temperatures.  In Fig.~\ref{fig:etatail} we plot the tail coefficient
$C_\eta$ extracted from our numerical results for the frequency
dependent shear viscosity as a function of temperature.  Apparently,
the connection 
\begin{align}
  \label{eq:Ceta}
  C_\eta = \frac{C}{15\pi}
\end{align}
between the tail coefficient $C_\eta(T)$ and the Tan contact density $C(T)$ 
that follows from the exact sum rule \eqref{eq:sumeta2} is fulfilled very accurately at 
all temperatures (for an alternative derivation of the relation \eqref{eq:Ceta}
see \ref{sec:contact_and_tail_coefficient}).  To understand the temperature 
dependence of the tail coefficient, which saturates for low temperatures and 
decreases as $\sim 1/T$ for large temperatures $T\gtrsim T_F$, we note 
that the contact density is defined via the asymptotics of the
momentum distribution, $n_\sigma(k) \to C/k^4$ for large wavenumbers
$k\gg k_F$.
Alternatively, the contact density can be determined from the vertex function
\eqref{eq:Cvtx} as $\hbar^4C=-m^2\Gamma_{X'=X^+}$
\cite{haussmann2007tbb}.  At high temperatures $T\gg T_F$ this can be
evaluated analytically and leads to 
\begin{align}
  \label{eq:Chightemp}
  C(T) = \frac{4m^2z^2T^2}{\pi\hbar^4} = \frac{8\pi^2\hbar^2n^2}{mT} =
  \frac{16k_F^4}{9\pi^2\theta} \, ,
\end{align}
which agrees precisely with the result obtained in
Ref.~\cite{yu2009short} (note the different definition of the contact
in this work which accounts for an apparent difference by a factor of
$4\pi^2$). 
This asymptotic behavior is in perfect agreement with our numerical
results in Fig.~\ref{fig:etatail}.  An alternative way to infer the
high frequency behavior of the shear viscosity is based on the
relation \cite{taylor2010viscosity}
\begin{align}
  \eta(\omega) = \lim_{q\to0}
  \frac{3\omega^3}{4\hbar q^4}\, \Im\chi_{\rho\rho}(\q,\omega) 
\end{align}
between the frequency-dependent shear viscosity and the
mass-density correlation function $\chi_{\rho\rho}(\q,\omega)$, 
a relation that is valid at all 
frequencies.  As shown by Son and Thompson \cite{son2010short}, the
density correlation function at large frequencies 
\begin{align}
  \Im\chi_{\rho\rho}(\q,\omega\to\infty)
  = \frac{4\hbar^{5/2}q^4C}{45\pi m^{1/2}\omega^{7/2}} 
\end{align}
is again fully determined by the Tan contact $C$.  The resulting
coefficient $C_\eta$ in the high-frequency tail of the shear viscosity
agrees precisely with that in equation \eqref{eq:Ceta} above.

\begin{figure}
  \includegraphics[angle=-90,width=.8\linewidth,clip]{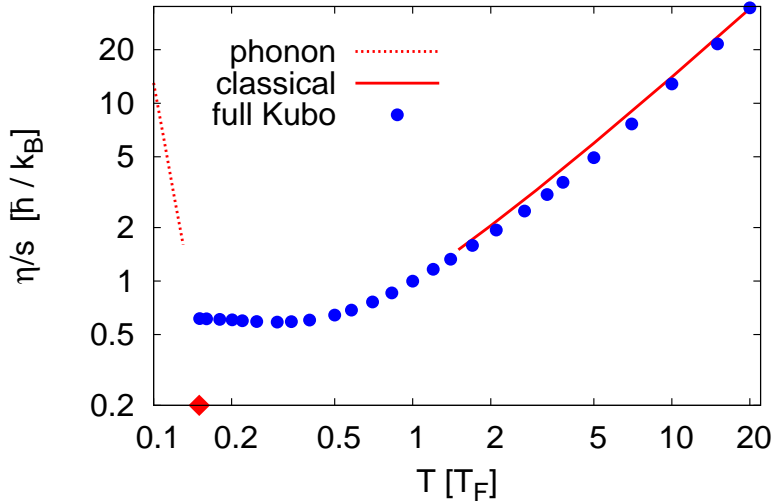}
  \caption{[color online] Shear viscosity to entropy ratio $\eta/s$
    (blue circles) in comparison with known asymptotes.  The dashed
    red line on the left is the phonon contribution $\eta/s \sim
    (T/T_F)^{-8}$ in equation \eqref{eq:landau_ufg}, the solid red
    line on the right the classical limit \eqref{eq:etaclass} divided
    by the classical entropy from the Sackur-Tetrode formula.  The red
    diamond indicates $T_c \simeq 0.15\,T_F$.}
  \label{fig:etas}
\end{figure}
Our result for the temperature-dependent shear viscosity can now be
combined with the known value of the entropy density
\cite{haussmann2007tbb} to determine the ratio $\eta/s$ in the normal
fluid regime of the unitary gas.  As shown in Fig.~\ref{fig:etas},
this ratio exhibits a very shallow minimum around $T\approx 0.3-0.4\,
T_F$, below which $\eta/s$ increases very slowly. The precise location
of the minimum clearly depends sensitively on how accurate the results
for both the viscosity and entropy are in this regime.  On quite
general grounds, it is likely that the minimum in $\eta/s$ is close to
the superfluid transition temperature $T_c\simeq 0.15\, T_F$, and that
$\eta/s$ is monotonically increasing as the temperature is lowered in
the superfluid regime, eventually crossing over to the steep increase
of equation \eqref{eq:landau_ufg} as the temperature approaches zero.
The fact that our diagrammatic calculation gives results, e.g., for
the critical temperature and the associated entropy density $s\simeq
0.7\, nk_B$ \cite{haussmann2007tbb} which agree well with precise
numerical results \cite{burovski2008critical}, suggests that our ratio
$\eta/s$ provides a quantitatively reliable estimate, despite the fact
that the precise location of the minimum is difficult to determine.
Granting that our value $\eta/s \simeq 0.6\, \hbar/k_B$ for the
minimum is close to the exact result, we conclude that the ratio
$\eta/s$ for the unitary Fermi gas remains a factor of about seven
above the KSS bound, somewhat larger than the experimental estimates
for the quark-gluon plasma~\cite{schaefer2009npf}.

%%%%%%%%%%%%%%%%%%%%%%%%%%%%%%%%%%%%%%%%%%%%%%%%%%%%%%%%%%%%%%%%%%%%%%%%

\section{Conclusions}
\label{sec:concl}

Using a diagrammatic approach that follows the classic Baym-Kadanoff
conserving approximation to calculate transport properties of
fermionic quantum liquids, we have determined the frequency dependent
shear viscosity $\eta(\omega)$ of the unitary Fermi gas in its normal
phase $\theta>\theta_c \approx 0.15$.  With decreasing temperatures,
the Drude peak in the viscosity is smeared out and spectral weight is
transferred to the $1/\sqrt{\omega}$ tail, precisely following the
increase in the Tan contact.  In the degenerate gas regime
$\theta\lesssim 1$, transport of transverse momentum cannot be
described in terms of a quasiparticle picture because the viscosity
spectral function is much broader than the thermal energy.  Our 
results for $\eta(\omega)$ are in perfect agreement with the exact
sum rule \eqref{eq:sumeta2} which, apart from numerical factors, 
agrees with the one derived by Taylor and Randeria \cite{taylor2010viscosity}.
In particular, the weight of the Drude peak, which is effectively the
fluid's high-frequency shear modulus, is identically equal to the gas
pressure at all temperatures above $T_c$, while the coefficient
$C_\eta = C/(15\pi)$ of the high-frequency $1/\sqrt{\omega}$ tail agrees
with the Tan contact density.

Our dynamic shear viscosity exhibits a $\sqrt\omega$ singularity
both at high and low frequencies.  While the high-frequency tail is a
consequence of the zero-range interaction, providing a direct measure
of the contact, the physical interpretation and a detailed theory of the
singularity at low frequencies that is related to the well-known
$t^{-3/2}$ long-time tails due to mode coupling effects remains an
open problem.  Note that within our formulation, the sign of the
$\sqrt\omega$ singularity is positive, as is found for quantum liquids
\cite{kirkpatrick2002long}, while the standard long-time tails in
classical liquids lead to a negative contribution. 

The proper treatment of 
symmetries and the inclusion of vertex corrections guarantees that our
diagrammatic approach leads to a bulk viscosity that vanishes
identically.  In particular, our approximation exactly satisfies the Tan
energy relation \eqref{eq:tanenergy} and the adiabatic sweep theorem
\cite{tan2008large}. 
Moreover, it reproduces the high-temperature Boltzmann
equation result for the shear viscosity by Bruun and Smith
\cite{bruun2007sva}.  The accuracy of our approximation in the most
interesting regime just above the superfluid transition is difficult
to estimate a priori.  The agreement with the
non-perturbative viscosity sum rule, however, suggests that our
results for the viscosity are quantitatively reliable down to the
superfluid transition.  The dc-viscosity is found to decrease
monotonically with temperature, reaching a value $\eta\simeq
0.5\,\hbar n$ near $T_c$.  Equivalently, the unitary gas near $T_c$
exhibits a quantum limited shear diffusion constant (or kinematic
viscosity) $D_\eta\simeq 0.5\,\hbar/m$.  The ratio $\eta/s$ between
shear viscosity and entropy density exhibits a minimum slightly above
the superfluid transition temperature.  The corresponding value
$\eta/s \simeq 0.6\, \hbar/k_B$ is a factor of about seven above the
KSS bound, the smallest value found for non-relativistic fluids so
far.
 
There are, of course, a number of problems that remain open.  For the
unitary gas, precise results for the shear viscosity near $T_c$ and
also immediately below the superfluid transition are required to
obtain definite predictions for the location and the associated value
of the minimum in both $\eta$ and in $\eta/s$.  Similarly, a
quantitative understanding of current experiments on the
inter-diffusion of the two spin components
\cite{zwierlein2010diffusion} requires to study particle transport by
a calculation of the conductivity from the longitudinal current
correlation function.  Another transport coefficient that may be
determined by an extension of our present method is the heat diffusion
constant.  Its ratio with the shear diffusion constant determines the
Prandtl number, which is equal to unity in non-relativistic, conformal
field theories with a holographic dual ~\cite{rangamani2009conformal}.

Finally, it is of interest to study the situation away from unitarity,
where scale invariance is broken and a finite bulk viscosity appears.
Since both $\eta$ and the ratio $\eta/s$ will exhibit a minimum as a
function of temperature for {\it arbitrary} values of $1/k_Fa$ along
the BCS-BEC crossover, this would allow to answer the question whether
the minimum found at unitarity indeed gives the lowest possible value
and thus, whether scale invariance is an important ingredient for a
minimum in $\eta/s$ in a non-relativistic context.  In fact, the
entropy density at $T_c$ has its largest value not at unitarity but
near $1/k_Fa \approx 1$ \cite{haussmann2007tbb}.  A calculation of the
viscosity along the complete BCS-BEC crossover has been performed in
one dimension, where a finite bulk viscosity exists at $T=0$
\cite{punk2006collective}. In this case, the zero temperature
viscosity is actually smallest in the BEC limit.

We are grateful for a number of fruitful discussions with Dietrich
Belitz, Jens Braun, Wolfgang G\"otze, John McGreevy, Yusuke Nishida,
Mohit Randeria, Achim Rosch, Manfred Salmhofer, J\"org Schmalian,
Richard Schmidt, Thomas Sch\"afer and Martin Zwierlein.  This work has
been supported in part by the DFG research unit ``Strong Correlations
in Multiflavor Ultracold Quantum Gases''.

%%%%%%%%%%%%%%%%%%%%%%%%%%%%%%%%%%%%%%%%%%%%%%%%%%%%%%%%%%%%%%%%%%%%%%%%

\appendix

\section{Viscosity of a Lennard-Jones fluid}
\label{sec:LJ}

In this Appendix we give an elementary argument, based on dimensional
analysis only, for the quite surprising observation that the KSS bound
which arises from genuine quantum field theory considerations, even
applies to classical fluids, whose viscosity minimum apparently cannot
depend on $\hbar$.  An example is water, a rather complex fluid, that
has a minimum viscosity near its critical point at $T=650\,$K.  Since
a viscosity has units energy density times seconds, the ratio
$\alpha_{\eta}=\eta/(\hbar n)$ is a pure number, which happens to be
of order one, $\alpha_{\eta}\simeq 32$ at the viscosity minimum. This
appears to indicate that quantum mechanics determines the minimum
viscosity of water despite the fact that the thermal wavelength
$\lambda_T$ near its critical point is much smaller than the average
interparticle spacing $n^{-1/3}$.

To understand the origin of this surprising fact, consider a simple
model for a classical fluid, which assumes pairwise central forces
that are derived from a Lennard-Jones potential
\begin{equation}
  \label{eq:Lennard-Jones}
 \Phi(r)=4\varepsilon\left[\left(\frac{\sigma}{r}\right)^{12}-
     \left(\frac{\sigma}{r}\right)^{6}\right] \, .
\end{equation}
This potential is, of course, not a realistic model for such complex
fluids as water, yet it provides a rather good quantitative
description of simple fluids like Argon \cite{hansen2006liquids}.  The
two parameters $\sigma$ and $\varepsilon$ of this potential provide
the characteristic scales for length and energy.  As a result, the
equation of state $p(n,T)=\varepsilon/\sigma^3\, p^{\star}(n\sigma^3,
T/\varepsilon)$ is completely determined by a universal function
$p^{\star}(n^{\star},T^{\star})$ of the dimensionless density
$n^{\star}=n\sigma^3$ and the dimensionless temperature
$T^{\star}=T/\varepsilon$.  This provides an explanation for the law
of corresponding states for different fluids that are reasonably well
described by an interaction of the Lennard-Jones type.  To evaluate
dynamic correlation functions like the viscosity from a (classical)
Kubo formula for this Lennard-Jones fluid requires in addition the
particle mass $m$, which fixes a characteristic time scale
$\tau=(m\sigma^2/\varepsilon)^{1/2}$.  By purely dimensional arguments
the shear viscosity of the Lennard-Jones fluid is then of the form
\begin{equation}
  \label{eq:LJ-viscosity}
 \eta_{\rm LJ}(n,T)=\frac{\varepsilon\tau}{\sigma^3}\;
 \eta^{\star}\left(n^{\star}, T^{\star}\right)
\end{equation}
with a universal function
$\eta^{\star}\left(n^{\star},T^{\star}\right)$.  In fact, equation
\eqref{eq:LJ-viscosity} also applies for the bulk viscosity $\zeta$,
which is typically of the same order as $\eta$ in classical fluids
\cite{hansen2006liquids}.  In the high-temperature, gaseous limit
$T^{\star}\gg 1$ this function can be calculated from kinetic theory,
where $\eta(T)\simeq\langle p\rangle/\sigma(T)$ is a ratio between an
average momentum and a thermally averaged cross section $\sigma(T)$.
For large temperatures, the cross section $\sigma(T)\sim p_T^{-4/n}$
scales with an inverse power of the thermal momentum $p_T$. The
exponent is determined by the short range repulsive potential
$\Phi(r\to 0)\sim r^{-n}$. For a Lennard-Jones fluid with $n=12$, the
universal function $\eta^{\star}\left(n^{\star}, T^{\star}\right)\sim
(T^{\star})^{2/3}$, therefore, exhibits a power-law increase with
temperature, independent of density.  At low temperatures the
viscosity will again increase and eventually the response
$\eta(\omega)\sim i/\omega$ becomes purely reactive at the
liquid-to-solid transition, which appears at densities
$n^{\star}\gtrsim 0.86$ \cite{hansen2006liquids} (Note that this is
the density at the freezing point, which is smaller than the melting
point density due to the first order nature of the transition. Note
also that a gas of hard spheres with diameter $\sigma$ reaches close
packing only at $n^{\star}=\sqrt{2}$).  Numerical results for the
dimensionless function $\eta^{\star}$ are available from
molecular-dynamics simulations.  For the Lennard-Jones fluid, they
were first performed by Levesque et al.\ \cite{levesque1973computer}.
For example, the resulting value is $\eta^{\star}\left(n^{\star}=0.72,
  T^{\star}=0.84\right)=3.29$ near the triple point and close to the
solidification line \cite{viscardy2007transport}. The minimum of the
viscosity is necessarily reached near the gas-liquid transition, which
lies between the triple and the critical point.  For the Lennard-Jones
fluid, the dimensionless densities and temperatures at these points
are of order unity: $n^{\star}_t\simeq 0.85,\;T^{\star}_t\simeq 0.68$
and $n^{\star}_c\simeq 0.36,\; T^{\star}_c\simeq 1.36$, respectively
\cite{hansen2006liquids}.  From equation \eqref{eq:LJ-viscosity}, it
is thus obvious that the minimum value of the shear viscosity of a
Lennard-Jones liquid
\begin{equation}
  \label{eq:LJ-viscosityminimum}
  \eta_{\rm LJ}^{\rm min}
  =\textit{const}\;\frac{\sqrt{m\varepsilon}}{\sigma^2}
\end{equation}
is fixed by the microscopic ratio $\sqrt{m\varepsilon}/\sigma^2$ with
a universal prefactor of order unity, that unfortunately seems not to
have been calculated so far.  Clearly, this minimum value does not
contain $\hbar$.  Indeed, the condition for the applicability of
classical physics to describe the behavior near the gas-liquid
transition of the Lennard-Jones fluid is $\lambda_T\ll
n^{-1/3}\simeq\sigma$, which implies that
$\sqrt{m\varepsilon}\gg\hbar/\sigma$.  The minimum value of the
viscosity is therefore much larger than $\hbar/\sigma^3\simeq\hbar n$
as long as the standard de Boer parameter
\begin{equation}
  \label{eq:deBoer}
  \Lambda^{\star}=\frac{\hbar}{\sigma\sqrt{m\varepsilon}}
\end{equation}
is very small. In practice, however, the de Boer parameter is still
appreciable unless one considers fluids of very heavy elements like
Xe. The associated conversion factor from equation
\eqref{eq:LJ-viscosityminimum} to a purely quantum limited viscosity
$\eta=\alpha_{\eta}\hbar n$ is therefore not that large.  Stated
differently, for most fluids the zero-point energy $\hbar^2/m\sigma^2$
associated with the typical length scale of interatomic potentials is
not much smaller than the well depth $\varepsilon$.  This is a
reflection of the fact that, after all, it is the combination of the
quantum mechanical van der Waals interaction at large separations and
the short range scale $\sigma$ that arises from the Pauli principle
which determines the ratio $\sqrt{m\varepsilon}/\sigma^2$.  It is,
therefore, quantum mechanics which eventually fixes the minimum value
of the viscosity of classical liquids to be of order $\hbar$ times the
liquid density.  With a corresponding entropy density $s\simeq k_Bn$,
this immediately yields a minimum value of $\eta/s$ that is not too
far above the KSS bound, despite the fact that both the viscosity and
the entropy may be inferred by completely classical considerations.

%%%%%%%%%%%%%%%%%%%%%%%%%%%%%%%%%%%%%%%%%%%%%%%%%%%%%%%%%%%%%%%%%%%%%%%%

\section{Scale invariance and Ward identities}
\label{sec:scale}

In this Appendix we show that scale invariance and the associated
continuity equation for the conserved dilatation current imply Ward
identities which relate the fermionic and bosonic \emph{bulk}
viscosity vertices to the respective single-particle Green's
functions.  As a consequence the bulk viscosity vanishes identically
in our approach as required by symmetry.  In particular this allows us
to check the consistency of the viscosity vertices that have been
computed explicitly.

The trace of the stress tensor defined in equations \eqref{eq:Tij0}
and \eqref{eq:Tell},
\begin{align}
  \label{eq:Tii0}
  \Pi_{ii}(\q=0,t) 
  = \Pi_{\ell=0}(t)
  = \sum_{\p\sigma} 2\varepsilon_p\, c^\dagger_{\p\sigma}
  c_{\p\sigma} 
  + \int d^3x \int d^3r\, 2V(\r)
  \; : \! n_\uparrow(\x) n_\downarrow(\x+\r) \! : \;\;
  = 2H
\end{align}
is twice the Hamiltonian in scale-invariant models
\cite{hagen1972scale, nishida2007ncf}.  The expectation value of this
operator equation yields immediately $3p=2\epsilon$
\cite{ho2004universal}, as shown explicitly in equation
\eqref{eq:tanenergy}.  The fermionic bulk viscosity response function
$\tilde T_{\ell=0}$ in equation \eqref{eq:Ttilde} can thus be written
in imaginary time $\tau$ as
\begin{align}
  \label{eq:Tbulkvev}
  \Tilde T_{\ell=0}(\tau X_1X_1') =
  2\left\langle \mathcal T H(\tau) \psi(X_1) \psi^\dagger(X_1')
  \right\rangle \,.
\end{align}
The effects of scale invariance are already incorporated in equation
\eqref{eq:Tii0}, and the evaluation of expectation values of the
Hamiltonian yields a time derivative \cite[equation
(28)]{baym1961conserv},
\begin{align}
  \label{eq:dtH}
  \frac{d}{d\tau} & \left\langle H(\tau) \right\rangle_U 
  = \int d^3r \sum_{X'}
  \Bigl[ U_{XX'} \, \frac{\partial}{\partial \tau}\, G_{X'X} +
  U_{X'X} \, \frac{\partial}{\partial \tau}\, G_{XX'} \Bigr] 
\end{align}
where $X=(\r,\tau)$, $X'=(\r',\tau')$ and the external bilocal field
$U_{XX'}$ couples to fermion bilinears via a term $\sum_{XX'}
\psi^\dagger(X)\, U_{XX'}\, \psi(X')$ in the action.  The coefficient
of equation \eqref{eq:dtH} linear in $U_{X_1'X_1}$ is then
\begin{align}
  \frac 12 \, \frac{d}{d\tau} \Tilde T_{\ell=0}(\tau X_1X_1') 
 \label{eq:dtTii}
  & = \frac{d}{d\tau} \left\langle H(\tau) \psi(X_1)
    \psi^\dagger(X_1') \right\rangle \\
  & = \hbar \delta(\tau-\tau_1') \frac{\partial}{\partial \tau_1'}\,
  G_{X_1X_1'}  
  +\hbar \delta(\tau-\tau_1) \frac{\partial}{\partial \tau_1}\,
  G_{X_1X_1'} \,.  \notag
\end{align}
A Fourier transform of the viscosity response function from imaginary
time $\tau$ to Matsubara frequency $i\omega_m$ and from $X$, $X'$ to
incoming momenta $K_1=(\k,i\epsilon_n)$ and outgoing momenta
$K_1'=(\k,i\epsilon_n+i\omega_m)$ yields $\Tilde T_{\ell=0}(i\omega_m
K_1'K_1)$, or
\begin{multline}
  \label{eq:WIferm}
  \frac{i\hbar\omega_m}{2}
  \Tilde T_{\ell=0}(k,i\omega_m,i\epsilon_n+i\omega_m,i\epsilon_n) \\
  = (i\hbar\epsilon_n+i\hbar\omega_m+\mu)\,
  G(k,i\epsilon_n+i\omega_m) 
  - (i\hbar\epsilon_n+\mu)\, G(k,i\epsilon_n) \,.
\end{multline}
Hence, the bulk viscosity response function is given by a difference
of two single-particle Green's functions multiplied by their energies.
This Ward identity holds exactly and has been derived without
reference to the T-matrix approximation used above in
section~\ref{sec:calc}.  One can easily check the Ward identity in the
non-interacting case where $G(k,i\epsilon_n) =
-(i\hbar\epsilon_n-\varepsilon_k+\mu)^{-1}$ and $\tilde
T_{\ell=0}^{(0)} (k,i\omega_m, i\epsilon_n+i\omega_m, i\epsilon_n) =
G(k,i\epsilon_n+i\omega_m)\, 2\varepsilon_k\, G(k,i\epsilon_n)$.  As
an alternative to starting with equation \eqref{eq:Tii0}, which is
already a consequence of scale invariance, we have also derived the
Ward identity by applying the local conformal transformation
$\beta(t)$ in Ref.~\cite{son2006general} to the generating functional
of the Green's functions, with the same result.  We note that scale
plus Galilean invariance imply conformal invariance
\cite{hagen1972scale} and that both are necessary for a vanishing bulk
viscosity \cite{son2007vbv}.

A corresponding Ward identity holds for the bosonic viscosity response
function \eqref{eq:Stilde}, which is given by the difference of two
vertex functions $\Gamma(K)=\Gamma(\k,i\Omega_m)$,
\begin{multline}
  \label{eq:WIbos}
  \frac{i\hbar\omega_m}{2}
  \Tilde S_{\ell=0}(k,i\omega_m, i\Omega_m+i\omega_m, i\Omega_m) \\
  = (i\hbar\Omega_m+i\hbar\omega_m+2\mu)\,
  \Gamma(k,i\Omega_m) 
  - (i\hbar\Omega_m+2\mu)\, \Gamma(k,i\Omega_m+i\omega_m) \,.  
\end{multline}
These Ward identities provide a solution to the T-matrix transport
equations \eqref{eq:C_Fermi_vertex}--\eqref{eq:visc_bos}, as one can
check by plugging in the Ward identities \eqref{eq:WIferm} and
\eqref{eq:WIbos} on the right-hand side.  Thus we have proven that
the T-matrix approximation conserves not only charge, momentum and
energy but also the dilatation current (scale invariance).

From the Ward identities of scale invariance it follows that the bulk
viscosity computed via equation \eqref{eq:suscept} vanishes
for any frequency, temperature or density,
\begin{align}
  \label{eq:zeta0}
  \Re\zeta(\omega) \equiv 0 \,.
\end{align}
Indeed, the fermionic Ward identity \eqref{eq:WIferm} for $\tilde
T_\ell(K_1,K)$ implies by a shift of the frequency that the first term
of the stress correlation function \eqref{eq:suscept} vanishes
identically.  The bosonic Ward identity \eqref{eq:WIbos} implies,
again by a shift of frequency, that
\begin{align}
  \frac{1}{\beta} \sum_{i\Omega_m} \Tilde S_{\ell=0}(k,i\omega_m,
  i\Omega_m+i\omega_m, i\Omega_m) 
  = \frac{1}{\beta} \sum_{i\Omega_m} 4\,\Gamma(k,i\Omega_m) \,,
\end{align}
such that also the second term of the correlation function
\eqref{eq:suscept} vanishes identically.

This result has been obtained previously in different ways
\cite{werner2006unitary, son2007vbv, nishida2007ncf}.  For instance,
it follows in a rather direct manner from a sum rule
\begin{align}
 \label{eq:sumzeta}
  \frac{1}{\pi} \int_0^\infty d\omega\, \zeta(\omega)
  = \frac{1}{72\pi ma^2} 
  \left(\frac{\partial C}{\partial a^{-1}} \right)_s
\end{align} 
for the bulk viscosity that has been derived by Taylor and Randeria
\cite{taylor2010viscosity}.  The fact that the derivative $\partial
C/\partial(1/a)$ is finite at unitarity and that $\zeta(\omega)\geq 0$
is a positive function, immediately implies that the bulk viscosity of
the unitary gas at $a^{-1}=0$ vanishes at arbitrary frequencies.

%%%%%%%%%%%%%%%%%%%%%%%%%%%%%%%%%%%%%%%%%%%%%%%%%%%%%%%%%%%%%%%%%%%%%%%%

\section{Relation between contact and tail coefficient}
\label{sec:contact_and_tail_coefficient}

One can compute the viscosity tail $C_\eta$ defined in equation
\eqref{eq:etatail},
\begin{align}
  \label{eq:Aetatail}
  \eta(\omega\to\infty) & = \frac{\hbar^{3/2}C_\eta}{\sqrt{m\omega}}
\end{align}
in at least four different ways: from the current correlation
function, from the stress tensor correlation function, from the
relation between the shear viscosity and the density correlation
function, and from the viscosity sum rule \eqref{eq:sumeta2}. In
this Appendix we will analytically compute $C_\eta$ at unitarity by
the first two methods, thus confirming equation \eqref{eq:Ceta}.  The
identical result is obtained by considering the high-frequency
behavior of the density correlation function, as has already
been shown in section~\ref{sec:results}, and from the sum
rule. 

The trace of the current correlation functions is in general a linear
combination of shear and bulk viscosities \cite{taylor2010viscosity}
\begin{align*}
  \chi_J(\q,\omega)
  & \equiv \left\langle [j_k(\q,\omega),j_k(-\q,-\omega)]
  \right\rangle \\
  & = \chi_L(\q,\omega) + 2\chi_T(\q,\omega) \\
  & = \frac{iq^2}{\omega} \left[ \frac{10}{3} \eta(\omega) +
    \zeta(\omega) \right] + \mathcal O(q^4) \,.
\end{align*}
At unitarity $\zeta(\omega)\equiv 0$, and the real shear viscosity is
given by
\begin{align}
  \eta(\omega) = \frac{3}{10} \lim_{q\to0} \frac{\omega}{q^2}\,
  \chi_J''(\q,\omega) \,.
\end{align}
In the asymptotic high-frequency limit $\omega\to\infty$ only the
diagrammatic contributions at first order in the pair propagator
contribute to the tail.  These are the self-energy (S), Maki-Thompson
(MT) and Aslamazov-Larkin (AL) diagrams in Fig.~\ref{fig:firstorder}
(see also \cite[Fig.~6]{son2010short}),
\begin{align}
  \label{eq:chiJcontrib}
  \chi_J(\q,\omega) = \chi_J^\SE(\q,\omega) + \chi_J^\MT(\q,\omega) +
  \chi_J^\AL(\q,\omega) \,.
\end{align}
In each of these diagrams the vertex function $\Gamma_{XX'}$ is
replaced by its value at $X'=X^+$,
\begin{align}
  \label{eq:Cvtx}
  -\tilde g^2 \langle \Bar\psi_B \psi_B \rangle 
  = \Gamma_{XX'}\rvert_{X'=X^+} 
  = \sum_\k \frac{1}{\beta} \sum_{i\Omega_m} \Gamma(\k,i\Omega_m)
  = -\frac{\hbar^4C}{m^2}
\end{align}
which defines the contact density $C$ \cite{tan2008energetics}.  
The remaining part of the diagram is evaluated at zero bosonic
wavenumber $\k=0$ and Matsubara frequency $\Omega_m=0$,
\begin{align}
  \Gamma(\k,i\Omega_m) \mapsto -\frac{\hbar^4C}{m^2}\, \delta(\k)\,
  \delta(\Omega_m) \,.
\end{align}
To this order in the pair propagator, the fermionic Green's functions
are bare $G_0(\p,i\epsilon_n) =
-(i\hbar\epsilon_n-\varepsilon_\p)^{-1}$ and have spectral weight only
on-shell.  Then to absorb an external high-energy perturbation with
frequency $\omega$, fermions are excited to very high momenta
$2\varepsilon_\p \approx \hbar\omega$, which are much larger than the
Fermi or thermal momentum scales.  Hence only few-body physics is
probed and it is sufficient to evaluate the diagrams in the vacuum
limit $\mu=0$ and set the Fermi functions to zero.

Specifically, we obtain for the self-energy contribution to the
current correlation function \eqref{eq:chiJcontrib} 
\begin{align*}
  \chi_J^\SE(\q,i\omega_m)
  & = \frac{\hbar^4C}{m^2} \, \sum_\sigma \sum_\p j_k(\p,\p+\q)\,
  j_k(\p+\q,\p) \\
  & \qquad \times
  \frac{1}{\beta} \sum_{i\epsilon_n} G^2(\p,i\epsilon_n)
  G(-\p,-i\epsilon_n) 
  G(\p+\q,i\epsilon_n+i\omega_m)
  + (-\q,-i\omega_m) \\
  & = \frac{\hbar^4C}{m^2} \, 2\sum_\p 
  \frac{p^2}{4\varepsilon_{\p_-}^2
    (i\hbar\omega_m-\varepsilon_{\p_-}-\varepsilon_{\p_+})} 
  + (-\q,-i\omega_m)
\end{align*}
with the current vertex insertion $\p$ squared and shifted momenta
$\p_\pm=\p\pm\q/2$.  Likewise, the Maki-Thompson contribution is
\begin{align*}
  \chi_J^\MT (\q,i\omega_m) 
  & = \frac{\hbar^4C}{m^2} \, \sum_\sigma \sum_\p j_k(\p,\p+\q)\,
  j_k(\p+\q,\p) \\
  & \quad \times \frac{1}{\beta} \sum_{i\epsilon_n} G(\p,i\epsilon_n)
  G(-\p,-i\epsilon_n) 
  G(\p+\q,i\epsilon_n+i\omega_m) 
  G(-\p-\q,-i\epsilon_n-i\omega_m) \\ 
  & = \frac{\hbar^4C}{m^2} \, 2\sum_\p 
  \frac{p^2}{4\varepsilon_{\p_-} \varepsilon_{\p_+}} 
  \left(
    \frac{1}{i\hbar\omega_m-\varepsilon_{\p_-}-\varepsilon_{\p_+}} 
    +\frac{1}{-i\hbar\omega_m-\varepsilon_{\p_-}-\varepsilon_{\p_+}}
  \right) .
\end{align*}
Expanding to order $q^2$ and performing the integrals yields the
retarded correlation function
\begin{align}
  \chi_J^\SE(\q,\omega+i0)+\chi_J^\MT(\q,\omega+i0)
  & = \frac{(i-1)\hbar^{3/2}Cq^2}{3\pi m^{1/2}\omega^{3/2}}
\end{align}
and the shear viscosity tail
\begin{align}
  \eta_{\SE+\MT}(\omega\to\infty) = \frac{\hbar^{3/2}C}{10\pi\sqrt{m\omega}} 
\end{align}
with tail coefficient $C^{\SE+\MT}_\eta = C/(10\pi)$.

In the Aslamazov-Larkin contribution, either one of the two vertex
functions is
replaced by the contact and the other one is left in place, with the
external frequency $\omega$ and momentum $\q$ as arguments:
\begin{align}
  \chi_J^\AL(\q,i\omega_m) = -\frac{\hbar^4C}{m^2}\,
  [S_{k,\ell=1}(\q,i\omega_m)]^2\, \Gamma(\q,i\omega_m)
  + (-\q,-i\omega_m) \,.
\end{align}
The $\ell=1$ bosonic response vertex with one current vertex insertion
$p_k$ is given by
\begin{align*}
  S_{k,\ell=1}(\q,i\omega_m)
  & = -2\sum_\p j_k(\p,\p+\q)\, \frac{1}{\beta} \sum_{i\epsilon_n}
  G(\p,i\epsilon_n) 
  G(-\p,-i\epsilon_n) G(\p+\q,i\epsilon_n+i\omega_m) \\
  & = 2\sum_\p \frac{p_k}{2\varepsilon_{\p_-}
    (i\hbar\omega_m-\varepsilon_{\p_-}-\varepsilon_{\p_+})} \,.
\end{align*}
Integration and analytical continuation to a retarded function yields
\begin{align*}
  S_{k,\ell=1}(\q,\omega+i0)
  & = \frac{im^{3/2} q_k}{6\pi\sqrt{\hbar\omega}} \,.
\end{align*}
Since $[S_{k,\ell=1}(\q,\omega)]^2 = -m^3q^2/(36\pi^2\hbar\omega)$ is
already of order $q^2$ it is sufficient to evaluate the vertex
function \eqref{eq:Gammaret} at zero momentum,
$\Gamma^\ret(\q=0,\omega) = -4\pi i \hbar^{5/2} m^{-3/2}
\omega^{-1/2}$ to obtain the AL correlation function
\begin{align*}
  \chi_J^\AL(\q,\omega+i0)
  & = -i\,\frac{\hbar^{3/2}Cq^2}{9\pi m^{1/2}\omega^{3/2}}
\end{align*}
and the shear viscosity tail
\begin{align}
  \eta_\AL(\omega\to\infty) = -\frac{\hbar^{3/2}C}{30\pi\sqrt{m\omega}} \,.
\end{align}
The resulting total tail is
\begin{align}
  \label{eq:Aetaomega}
  \eta(\omega\to\infty) = \frac{\hbar^{3/2}C}{15\pi\sqrt{m\omega}} 
\end{align}
with tail coefficient $C_\eta = C/(15\pi)$, in agreement with both the
exact sum rule \eqref{eq:sumeta2} and our numerical data.

An alternative way to compute the tail in the frequency dependent
viscosities is to consider the stress tensor correlation functions in
the limit $\omega\to\infty$.  The calculation is analogous to the case
of the current correlations, except that there are now stress vertex
insertions, and one can let the external momentum $\q\to0$.  Note that
the interaction contribution to the shear stress operator vanishes for
the zero-range model of the unitary Fermi gas, cf.\
section~\ref{sec:calc}.  We obtain
\begin{align}
  C_\eta^{\Pi,\SE} & = C_\eta^{\Pi,\MT} = \frac{C}{30\pi}, &
  C_\eta^{\Pi,\AL} & = 0
\end{align}
in agreement with equation \eqref{eq:Aetaomega}.  For the case of the
bulk viscosity,
\begin{align}
  C_\zeta^{\Pi,\SE} & = C_\zeta^{\Pi,\MT} = \frac{C}{18\pi}, &
  C_\zeta^{\Pi,\AL} & = -\frac{C}{9\pi}
\end{align}
such that the total $C_\zeta = 0$ vanishes in accordance with scale
invariance \eqref{eq:zeta0}.

\end{document}